\documentclass[twoside,12pt]{article}
\usepackage{epsfig}
\usepackage{lineno}
\usepackage{cite}

\newcommand{\be}{\begin{equation}}
\newcommand{\ee}{\end{equation}}
\newcommand{\bea}{\begin{eqnarray}}
\newcommand{\eea}{\end{eqnarray}}

\begin{document}

\title{ \vspace{1cm} Nucleosynthesis in Type I X-ray
 Bursts}
\author{A. Parikh$^{1,2,*}$, J. Jos\'e$^{1,2}$, G. Sala$^{1,2}$, C. Iliadis$^{3,4}$ \\
\\
\small $^1$Departament de F\'isica i Enginyeria Nuclear, \\ \small Universitat
Polit\`ecnica de Catalunya (EUETIB), E-08036 Barcelona, Spain\\
\small $^2$Institut d'Estudis Espacials de Catalunya (IEEC), E-08034 Barcelona,
Spain\\
\small $^3$Department of Physics and Astronomy, University of North Carolina at
Chapel Hill, \\ \small Chapel Hill, NC 27599, USA\\
\small $^4$Triangle Universities Nuclear Laboratory, Duke University, Durham, NC
27708, USA\\}
\maketitle

\begin{abstract} Type I X-ray bursts are thermonuclear explosions that
occur in the envelopes of accreting neutron stars. Detailed
observations of these phenomena have prompted numerous studies in theoretical
astrophysics and
experimental nuclear physics since their discovery over 35 years ago. In this
review, we begin by discussing key observational features of these phenomena
that may be sensitive to the particular patterns of nucleosynthesis from the
associated
thermonuclear burning. We then summarize efforts to model type I X-ray bursts,
with emphasis on determining the nuclear physics processes involved throughout
these bursts. We discuss and evaluate limitations in the models, particularly
with
regard to key uncertainties in the nuclear physics input. Finally, we examine
recent, relevant experimental measurements and outline future prospects to
improve our understanding of these unique environments from observational,
theoretical and
experimental perspectives.
\end{abstract}


\vspace{55 mm}
{\footnotesize *Corresponding author.  Email address: anuj.r.parikh@upc.edu (A. Parikh). }

\newpage
\tableofcontents

\section{Introduction}
\label{intro}

Type I X-ray bursts are the most frequent types of thermonuclear stellar
explosions in the Galaxy. As well, they rank third after supernovae and classical novae in terms of total energy output in a single event, for phenomena observed in the Galaxy. Ninety-six
Galactic low-mass X-ray binary
systems exhibiting such bursting behaviour have been discovered to date\footnote{See http://www.sron.nl/$\sim$jeanz/bursterlist.html for a list of
known Galactic bursting sources.}. 

X-ray bursts were discovered in 1975 \cite{gri76} through observations made with
the Astronomische Nederlandse Satelliet \cite{ANS} of a previously known X-ray source, 4U
1820-30\footnote {Most X-ray sources are named using letters from the satellites
that
discovered them (e.g., 4U stands for the fourth catalogue of the satellite Uhuru, the first X-ray observatory) and numbers corresponding to their
coordinates in right ascension (1820 stands for 18h20min) and declination
(-30 deg)
in the sky. These sources may also be named after the constellation where they
are located and the order of discovery: 4U 1820-30 is also known as
Sgr X-4, the fourth X-ray source discovered in the constellation Sagittarius. To complicate matters further, even within one of these naming conventions a single source may have two names.  For example, the source 4U 1820-30 has been referred to in the literature as 3U 1820-30 as well.}. While a similar event had been observed in 1969 from the source Cen X-4 using the Vela 5B satellite \cite{Vela,bel72}, the authors did not relate the observed features to a new, distinct type of emission.  Consequently, that source was not associated with what we now call type I X-ray bursts until 1976 \cite{Bel76,kul09}.  At a distance of $\approx$1~kpc \cite{kul09}, Cen X-4 is the nearest-known X-ray bursting source \cite{int10} and has yielded the brightest X-ray burst ever recorded ($\approx$60~Crab\footnote{1
Crab = 2.4$\times10^{-8}$ erg s$^{-1}$ cm$^{-2}$ in the 2--10 keV band.}\cite{kul09}). These
initial, pioneering discoveries were soon followed by the identification of more
bursters. While most X-ray satellites have detected X-ray bursts, great advances in the field were made through observations by EXOSAT (operating from 1983 to 1986\cite{EXOSAT}), BeppoSAX (operatiing from 1996 to 2003\cite{BeppoSAX}), and the Rossi X-ray Timing Explorer (RXTE, operating from 1995 to 2012 \cite{RXTE}).  RXTE, with its large collecting area and capability for high temporal resolution X-ray analysis, has observed more than 1000 bursts from about 50 sources \cite{gal08}.

Soon after the discovery of X-ray bursting sources, it was
proposed that the explosions could be powered by thermonuclear runaways on the surface of 
accreting neutron stars \cite{WT76,MC77} (see Section \ref{mech}).  The discovery of the Rapid Burster in 1976 \cite{Lew76}, with burst recurrence times as short as 10 s, complicated matters.  The quick succession of flashes observed from this source seemed incompatible with the proposed mechanism, and moreover, did not match the general pattern shown 
by the other bursting sources.  A classification of type I and type II bursts was 
therefore established \cite{Hof78}, the former associated with thermonuclear flashes and
the latter linked to accretion instabilities. The Rapid Burster is one of only
two known sources showing type II bursts, together with the Bursting Pulsar (GRO 1744-28); furthermore, the Rapid Burster is the only source giving both type I and type II X-ray bursts\cite{Lew95}.
Type I bursts display
a spectral softening during the burst tail, reflecting a decrease in the effective temperature and thus the cooling 
of the neutron star atmosphere. Typically, light curve rise times range between $\approx$1 and $\approx$10 seconds and decay times range between $\approx$10 seconds and several minutes. Recurrence times may vary between roughly one and several hours.  In contrast,
type II bursts exhibit a constant temperature for the duration of the burst, which can range from $\approx$2 s to $\approx$700 s, with intervals between bursts from $\approx$7 s to $\approx$1 hr.  

More recently, other varieties of X-ray bursts of longer duration have 
been discovered. Superbursts, first identified from the source 4U 1735-444 by Cornelisse et al. (2000) \cite{Cor00,Kuu04}, have been observed in about 10\% of all bursters.  They are characterized by light curves with exponential-like decay 
times of several hours and include an extreme case, KS 1731--260, that lasted for about 12 hours \cite{Kuu02}. Roughly 1000 times more energetic than a typical X-ray burst (superbursts release $\approx$ 10$^{42}$ 
erg per burst), superbursts have much longer recurrence times of the order of a year.  The mechanism for these phenomena is thought to involve the unstable burning of carbon \cite{Cum01, Kee11, Kee12}.   As well, bursts intermediate in both energy (releasing $\approx$ 10$^{41}$ erg) and duration ($\approx$ 30 min) to typical type I X-ray bursts and superbursts have been observed \cite{intZand02,intZand05}.  These are thought to result from thermonuclear ignition within a layer of pure helium \cite{Cum06}, although other explanations may also be viable \cite{Fal09}.  We will not discuss these bursts of longer duration or type II bursts further. 

X-ray bursts (XRBs) occur on the surface of a neutron star belonging to a low-mass
X-ray binary (LMXB). LMXBs are binary systems where a faint, low mass (M $<$ 1
M$_\odot$) main
sequence or red giant star transfers material onto a neutron
star or black hole. The transferred matter is typically enriched in H, He or both H and He. Furthermore,
these mass transfer episodes are driven by Roche-lobe overflows, leading 
to the buildup of an accretion disk that surrounds the compact
object.  (In contrast to LMXBs, high-mass X-ray binaries consist
of a compact object capturing a fraction of the stellar wind from a high-mass
companion star.) The
maximum mass accretion rate is assumed to be set by the Eddington limit\footnote{The Eddington
limit for the luminosity $L_{Edd}$ corresponds to the maximum luminosity
emitted by a stellar object for which the radiation pressure balances the
gravitational
attraction. For higher luminosities, the energy exceeding the Eddington limit
generates a radiation-driven wind. In the case of accreting sources, the Eddington
limit for the luminosity also puts a limit on the accretion rate $\dot
M_{Edd}$.}, which for accretion of H-rich material onto a 1.4 M$_\odot$ neutron star is $\dot M_{Edd} \approx 2 \times 10^{-8}$ M$_\odot$ yr$^{-1}$. XRB sources often have short orbital
periods, in the range of 0.2 -- 15 hr; notable
exceptions include Cir X-1, GX 13+1 and Cyg X-2, with $P_{orb}$ = 398.4 hr,
592.8 hr and 236.2 hr, respectively.

As accretion proceeds onto the surface of the neutron star, the transferred
material is gradually compressed to large densities and becomes degenerate. This
compression also heats the envelope, creating conditions favourable to the
ignition of the fuel through nuclear reactions. Since the envelope is initially degenerate, it cannot expand to accommodate the energy released through the nuclear processes. As a result, both the temperature and nuclear energy generation rate in the envelope increase.  This drives a thermonuclear runaway which results in an XRB. Note that the accreted envelope above the neutron star is only weakly degenerate; that is, the degeneracy is lifted as the temperature increases at early times. For example, for a neutron star accreting fuel with $Z/A$=0.5, degeneracy is lifted ($E_{Fermi} \leq E_{thermal}$ $\approx$ $kT$) as the temperature exceeds roughly 0.3 GK. However, the extreme ejection velocities required for material to escape from a neutron star are never achieved by the moderate amount of energy released by nuclear reactions.  Consequently neither appreciable mass ejection nor substantial envelope expansion results.  

Between individual bursts, the bulk of the observed, accretion-powered, X-ray flux arises from
the hottest and innermost regions of the accretion disk.   This persistent emission is
not very sensitive to the nature of the compact object. Hence, the detection of thermonuclear X-ray bursts from a LMXB is
actually the most reliable indicator of the presence of a neutron star as
opposed to a black
hole.

Owing to the novel associated nucleosynthesis patterns, as well as the corresponding interest of experimentalists in measuring the different interactions involved, we will now focus on mixed H/He accretors throughout (see e.g., Refs. \cite{intZand05,Gal06} for sources exhibiting bursts in He-rich environments).  During the
thermonuclear
runaway, a H/He-rich envelope is transformed to matter strongly enriched in
heavier nuclei via the $\alpha$$p$-process and the $rp$-process. These two processes
involve proton- or $\alpha$-particle induced reactions on stable and radioactive nuclei, interspersed by occasional $\beta$-decays, depending on the nuclear
structure of the nuclide reached by the abundance flow. Since the peak
temperatures reached during the thermonuclear runaway may exceed 1 GK,
many different
nuclear interactions are likely to occur during the XRB.  

In this review we begin by discussing observable features of type I
XRBs, highlighting characteristics in these phenomena that may depend upon
nucleosynthesis. We then focus upon treatments that have been
used to
model XRBs and examine how the predictions of these calculations compare to one
another and to observations. Next, we explain the roles of both theoretical
and experimental nuclear physics in models of XRBs, and discuss nuclear physics
uncertainties that either have been or remain to be addressed to better
constrain model predictions. We conclude with suggestions to guide
future observational, theoretical and experimental nuclear physics efforts to clarify
nucleosynthesis in type I
XRBs.

\section{Observational features of type I X-ray bursts}
\subsection{What do XRBs look like?}
\label{obsover}

Most LMXBs are transient X-ray sources, indicating that the rate of accretion from the
disk onto the compact object is not constant.  LMXBs are typically discovered
during periods of enhanced accretion rate, with persistent luminosities in the
range
0.001--0.1 $L_{Edd}$.  These periods are known as outbursts, and each may last for a few weeks
or months. After an outburst the source returns to quiescence,
disappearing from the X-ray sky for several months or decades. Type I X-ray
bursts typically
occur during outbursts, when the accretion rate is high enough to supply
sufficient fuel for a thermonuclear runaway.  Note, however, that some XRBs have been detected from LMXBs in apparent quiescence \cite{kul09}.

\begin{figure}[ht]
\begin{center}
\includegraphics[scale=0.35]{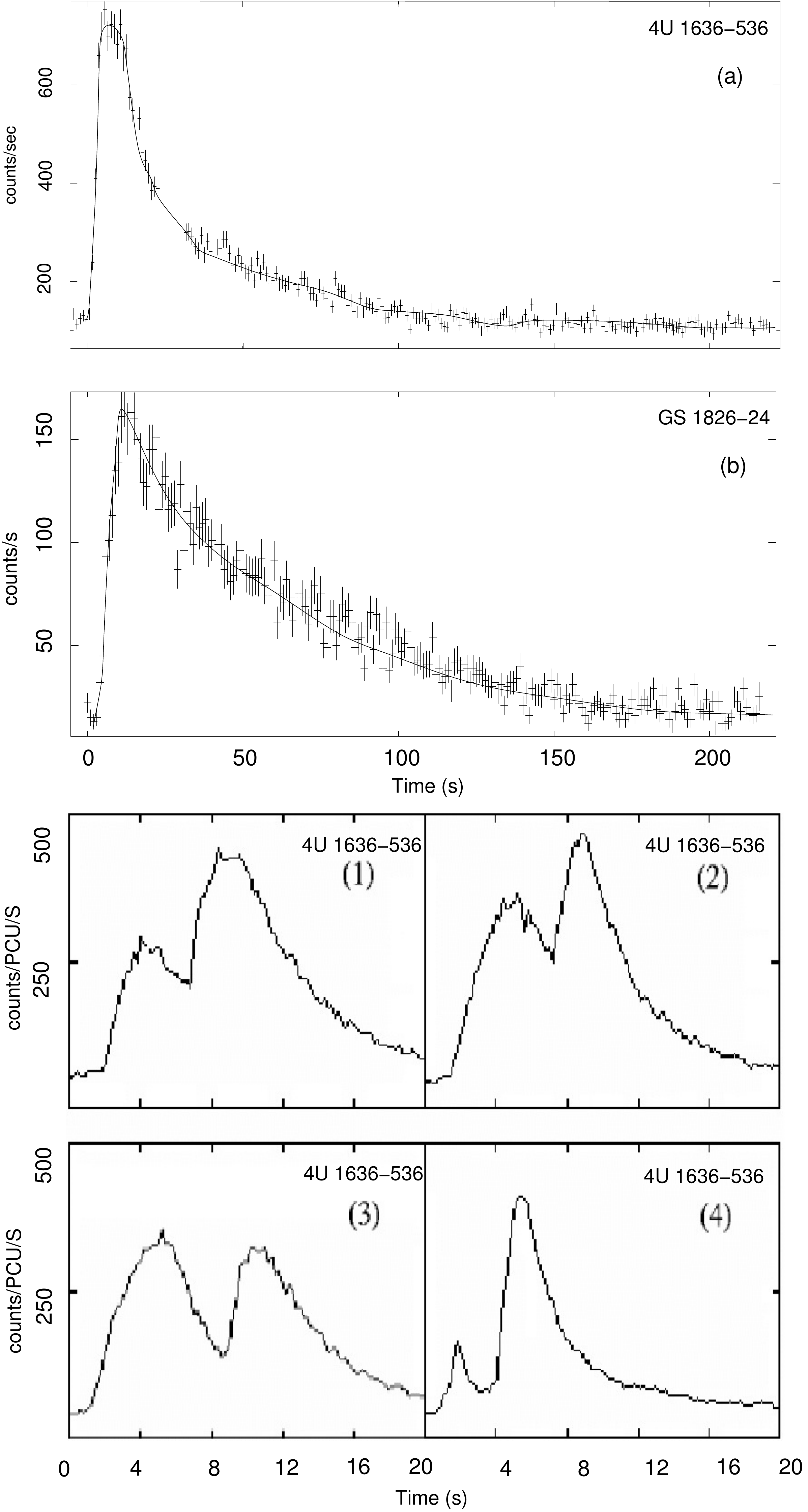}
\caption{Type I X-ray burst light curves from (a) 4U 1636-536, as observed with XMM-Newton/EPIC-pn \cite{XMM}, and (b) GS 1826-24, as observed with Swift/XRT \cite{Swift}. Typical rise times and decay times vary between $\approx$1 -- 10 s and $\approx$10 s -- several minutes, respectively.  Panels (a) and (b) were generated using public data, available at http://heasarc.nasa.gov/.  Panels (1--4) show four bolometrically double-peaked light curves from 4U 1636-536, as observed with RXTE \cite{RXTE} (PCU refers to RXTE Proportional Counter Units).  The separation of the two peaks is $\approx$3--5 s.  Credit for panels (1--4): A.L. Watts, I. Maurer, A\&A 467, L33, 2007, reproduced with permission, \copyright ESO.}
\label{lcs}
\end{center}
\end{figure}

Type I X-ray bursts appear as short flashes on
top of the persistent, accretion-powered emission.  Examples of light curves are shown in Fig.~\ref{lcs}.  Depending upon the source, bursts may occur either quite regularly (the most regular burster being GS 1826-24) or follow a highly variable and irregular pattern. If the burst recurrence time is well determined, the ratio of the integrated energy released between X-ray bursts to the integrated energy released during the burst defines the dimensionless $\alpha$ parameter. Assuming that all accreted fuel is burned during the burst, $\alpha$ should then correspond to the ratio of energies, per unit mass, liberated via accretion and thermonuclear burning.   Indeed, observed ratios of time-integrated
persistent and burst fluxes ($\alpha \approx40 - 100$ \cite{gal08}) match estimates based on the gravitational potential energy released
by matter falling onto a neutron star during accretion ($\approx 200$ MeV per nucleon) and the nuclear energy released during a burst ($\approx 5$ MeV per
nucleon, for fuel of solar composition).  These observations are also in agreement with detailed model calculations (see Section \ref{modresults}).  

In general, observations of X-ray bursts use detectors with medium to high time resolution.  For example, the PCA instrument onboard RXTE could measure light curves with a time resolution of 1 $\mu$s over an energy range of 2--60 keV, while the EPIC-pn instrument onboard the XMM-Newton observatory \cite{XMM} has a time resolution of 30 $\mu$s over an energy range of 0.1--10 keV.  These detectors also have medium to high spectral resolution, from $E/\Delta E\approx20-50$ for imaging cameras like EPIC, up to $E/\Delta E\approx200-800$ for spectrometers.  This
makes it possible to obtain time-resolved spectroscopy of X-ray bursts,
which shows that during the entire burst the source spectral energy distribution is usually well
approximated
by a blackbody spectrum (but see Section \ref{PRE}). From the blackbody fit to
the observed spectra, the colour (or blackbody) temperature $T_{bbody}$ and the bolometric flux
$F_{bol}$ (rate of energy emission per unit area over the whole energy
spectrum) are
obtained. If one assumes isotropic emission from a spherical surface, the radius $R_{NS}$ of the emitting object (as observed at infinity) can then be determined using

\begin{equation}
  F_{bol}=\frac{\sigma T^4_{bbody}R^2_{NS}}{f^4_{c}D^2} \left( 1+z \right)^{2} 
\end{equation}

\noindent where $\sigma$ is the Stefan-Boltzmann constant, $D$ is the distance from the 
observer to the X-ray burster, $M_{NS}$ is the mass of the neutron star and $f_{c}$ is the colour correction factor (see Section \ref{PRE}).    The gravitational redshift $z$ can be expressed as
$1+z=(1-2GM_{NS}/R_{NS}c^2)^{-1/2}$, which for a typical neutron star of
$M_{NS}=1.4$ $M_{\odot}$ and $R_{NS}=10$ km is $(1+z)=1.31$.  Many X-ray bursters are located in globular clusters with
well-determined
distances.
Measurements of radii of the emitting surface of X-ray bursters indicate values
around 10~km\cite{Lew95}.

Studies of the blackbody temperature evolution during XRBs
show an initial increase in temperature during the increase in flux from the
burst, namely, when thermonuclear burning occurs.  This is followed by the cooling of the
photosphere
during the tail of the burst.  This evolution occurs with a constant emitting area, that is, without significant expansion of the neutron star envelope (except in the
particular cases with photospheric radius expansion, see Section \ref{PRE}).

In general, the persistent X-ray flux $F_p$ between X-ray bursts is taken as an
indicator of the accretion rate $\dot M$, as \cite{gal08}
\begin{equation}
\dot
M=1.33\times10^{-11}\left(\frac{F_pc_{bol}}{10^{-9}\,erg\,cm^{-2}\,s^{-1}}\right)\left(\frac{D}{10\,kpc}\right)^2\left(\frac{M_{NS}}{1.4\,M_{\odot}}\right)^{-1}\left(\frac{1+z}{1.31}\right)\left(\frac{R_{NS}}{10\,km}\right)M_{\odot}\,yr^{-1}.
\end{equation}

\noindent The quantity $c_{bol}$ is a correction that must be made to convert the detected X-ray flux $F_p$ to the bolometric flux $F_{bol}$. 

Since X-ray bursts are powered by the fuel accumulated between bursts, it is
expected that the burst rate increases with increasing persistent flux, i.e.,
if material piles up faster, less time is needed to accumulate sufficient fuel for a
burst.
Indeed the behaviour of some bursters is in accord with this expectation (see e.g., 
Fig. 1 in Ref. \cite{lin12}). However, in several cases, the recurrence rate instead decreases
as the persistent emission increases \cite{gal08}. This may indicate that He is
burning
between bursts \cite{nar03,bil95,heg07}, that the local accretion rate is
actually decreasing in spite of an increase in the total accretion rate \cite{Bil00}, or that the persistent emission is in fact not a reliable
indicator of the accretion
rate.

Observational characteristics of X-ray bursting sources, e.g., time between
bursts, rise time of the light curve, burst duration, or total energy emitted,
depend on various parameters such as the accretion rate, the fuel composition
or the
neutron star spin. For accretion of solar metallicity material, Refs. \cite{Fuj81,FL87} identified several ignition regimes as a function of the adopted
mass-accretion rate $\dot M$:
\begin{itemize}
\item $\dot M < 2 \times 10^{-10}$ M$_\odot$ yr$^{-1}$: Simultaneous H- and He-burning triggered by thermally unstable H-ignition.  Here, the temperature in
the accreted layer is too low for stable hydrogen burning between bursts. H
ignites unstably in this case, triggering He burning and giving rise to a
type I
X-ray burst in a H-rich environment.
\item $2 \times 10^{-10}$ M$_\odot$ yr$^{-1}$ $< \dot M <$
$(4.4 - 11.1) \times 10^{-10}$ M$_\odot$ yr$^{-1}$: Pure He-burning following completion of H-burning.  Here, the temperature in the
accreted envelope is higher and H burns steadily to He between X-ray bursts. The
bursts occur as a consequence of He ignition in a pure He layer, resulting in
fast
and bright type I X-ray bursts.
\item $\dot M > (4.4 - 11.1) \times 
10^{-10}$ M$_\odot$ yr$^{-1}$: Simultaneous H- and He-burning triggered by thermally unstable 
      He-ignition.  Here, accretion provides fresh H fuel at a higher
rate than it is steadily burned. He ignites in a mixed H/He environment,
resulting in type I X-ray bursts of longer duration.
\end{itemize}

For sufficiently high accretion rates ($\dot M > 3.3 \times 10^{-8}$ M$_\odot$ yr$^{-1}$) stable burning is expected (for fuel of solar composition\cite{Fis07}).  The exact mass-accretion rate boundaries for the different regimes are
still under investigation through detailed models. For example, for accretion of
metal-deficient matter, lower values for the mass-accretion rates at the
transition
between the different ignition regimes have been obtained, narrowing the
predicted regime for pure He-burning \cite{Bil00}.  As well, recent simulations\cite{Fis07} predict that unstable burning persists for accretion rates higher than those inferred from observational constraints, calling into question other assumptions in the models (e.g., the composition of the accreted material).     

Since one source may evolve through different accretion rates during an
outburst, in principle the different burning regimes could be observed in the
same source. In several bursters, the transitions between burning regimes have
nicely shown up
through the observation of changes in burst durations\cite{chen07,zhang11,lin12}, but exceptions are abundant. Many
sources show the exact opposite behaviour from what is expected: burst durations decrease as
the apparent accretion rate increases for sources thought to be passing from the second to the third regime above\cite{mur80,lew87,lan87,mun00,corn03}.

\subsection{Photospheric radius expansion (PRE) bursts}
\label{PRE}

\begin{figure}[ht]
\begin{center}
\begin{minipage}[t]{8 cm}
\epsfig{file=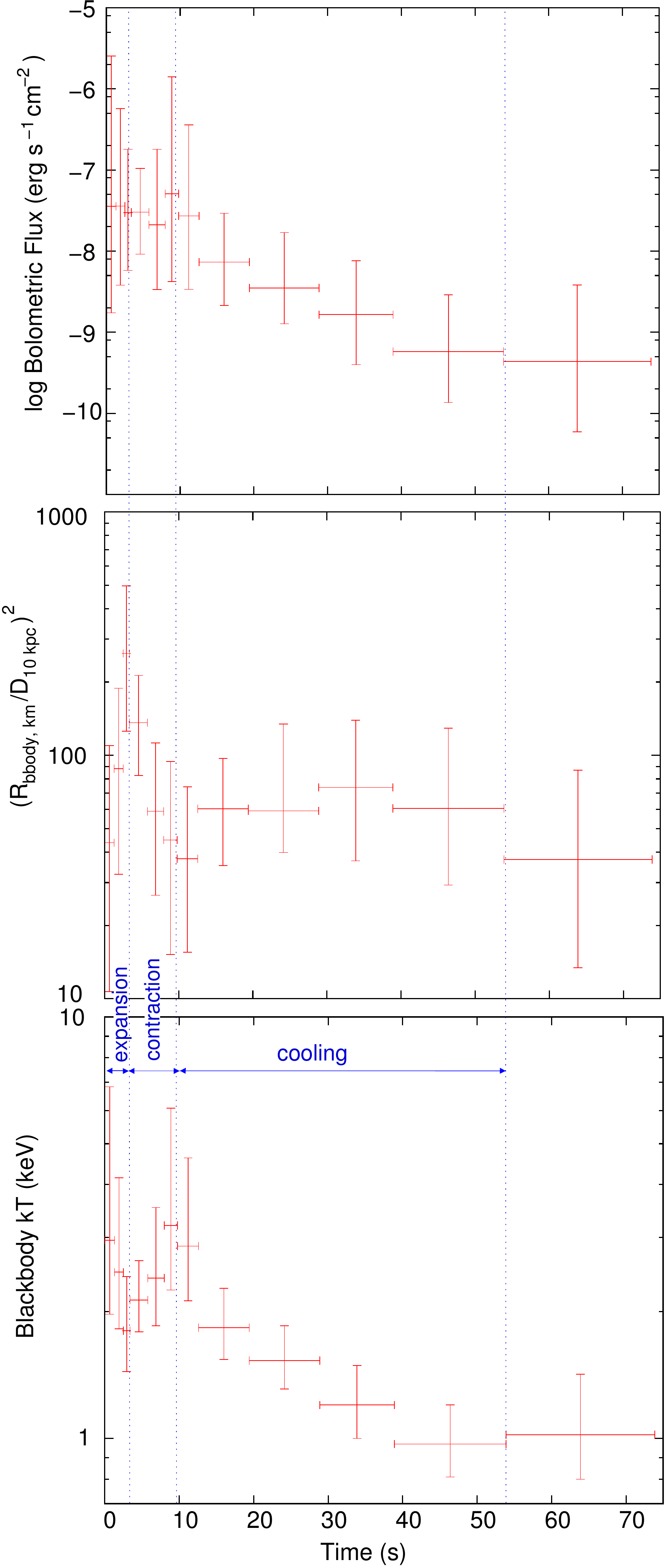,scale=0.5}
\end{minipage}
\begin{minipage}[t]{16.5 cm}
\caption{Evolution of the bolometric flux (top), emitting area (middle), and blackbody temperature T (bottom) during a type I X-ray burst of the Rapid Burster (MXB 1730-335) showing photospheric radius expansion.  More precisely, the ordinate of the middle panel shows a quantity proportional to the emitting area, where $R_{bbody}$ is the radius of the source (in km) and D is the distance to the source in units of 10~kpc.  Three phases are evident: expansion of the photosphere, lasting $\approx 3$ seconds; contraction of the photosphere, lasting $\approx 7$ seconds; and the final cooling of the envelope.  Data obtained with the Swift X-ray Telescope \cite{Swift}.  See Ref. \cite{sala12} for more details.}
\label{RB}
\end{minipage}
\end{center}
\end{figure}

In some bright type I X-ray bursts, the luminosity reaches the Eddington limit
and causes the expansion of the envelope. This is illustrated in Fig.~\ref{RB} for a type I burst observed from the Rapid Burster \cite{sala12}.  A first indicator of the expansion of the envelope is a
flat-topped light-curve, with a nearly constant bolometric luminosity
during the
envelope expansion (top panel). Confirmation of the expansion is provided by
time-resolved spectroscopy of the constant bolometric luminosity interval,
showing the increase (expansion) and decrease (contraction) of the photospheric
radius (middle panel) in parallel
with the decrease and increase of the effective temperature (bottom panel). This is
followed by the final cooling of the neutron star envelope.  See also, e.g., Fig. 3.4 from Ref. \cite{sb06} for examples of both PRE and non-PRE bursts from a single source.

In
some extreme cases, the temperature drop is large enough to shift the X-ray emission
to energies
below the range of sensitivity for typical instruments, causing an apparent
double-peak in the detected X-ray flux.  This effect should not be confused with intrinsic double
or multiple peaks of the bolometric flux (see Section~\ref{multipeak}).

Observations of these PRE bursts probe the still-unknown equation of state (EOS) of the neutron star through the determination of the
neutron star's mass and radius.  Several models for the EOS have been developed, 
with and without condensates, and some including strange quark matter \cite{Lat01,Lat07,Ste10}.
As each EOS predicts a particular relation between the mass and radius of the neutron star, these EOS models can be tested using observational constraints on masses and radii.  For a PRE burst located at a known distance, the neutron star mass can be determined from the observed
bolometric luminosity if one assumes that the maximum observed luminosity corresponds to the
Eddington limit. The neutron star radius can then
be
inferred from spectral fits during the cooling tail of the burst if one assumes the emitting
area to be the entire neutron star surface \cite{damen90,Ste10,guver11a,guver11b}.

A number of details play a role in the determination of the neutron star radius and mass from PRE bursts, and the composition of the envelope is a critical element in these calculations. The hydrogen mass fraction X is needed to determine the electron scattering opacity $\kappa_{e}=0.2(1+X)$ cm$^{2}$g$^{-1}$, which in turn is used to determine the Eddington limit for the luminosity (and hence the neutron star mass).  In addition, the composition of the atmosphere determines the colour correction factor: the neutron star spectrum is
known to differ from a perfect blackbody.  This is because electron scattering in the hot neutron star atmosphere suppresses part of the emission \cite{vanparadijs}.  The colour correction factor $f_c=T_{bbody}/T_{eff}$ relates the effective temperature $T_{eff}$ to the colour or blackbody temperature $T_{bbody}$ obtained from the
spectral fit. The quantity $f_c$ affects the determination of the emitting area, and hence the neutron star radius, since the emitting area is obtained from the observed flux and temperature. Detailed calculations using atmosphere models \cite{madej,majczyna,suleimanov} have determined colour correction factors for metallicities up to solar; the smallest colour correction factors were obtained assuming solar abundances because the absorption edges of Fe move the neutron star spectrum closer to a blackbody spectrum. However, in Eddington-limited
bursts, the photosphere of the neutron star is expected to contain high-metallicity ashes from the thermonuclear burning of H and He deep in the envelope \cite{wei06,intzand10}. This is in agreement with results from hydrodynamic simulations, which show that convective mixing
approaches the photosphere when the flux reaches the Eddington limit \cite{Woo04,Fis08,Jos10}. Increased atmospheric metallicity would imply more absorption features in the final emerging spectrum and even smaller colour correction factors.  Atmosphere model calculations adopting metallicities larger than solar are therefore urgently required to examine the impact of smaller colour correction factors on neutron star mass and radius determinations using PRE bursts. 

In addition to the uncertainties introduced by the differences between an ideal blackbody spectrum and the actual emission from the neutron star atmosphere during a XRB, the final observed spectrum may be altered by the interaction of the emerging spectrum
with the accretion flow or the hot corona surrounding the neutron star (e.g., through reflection, partial occultation, Comptonization). The resulting uncertainties seem to be sufficiently small \cite{guver11a,guver11b} to still allow one to
determine neutron star masses and radii with sufficient precision to distinguish between different neutron star EOS models.

There are therefore two ways in which the ashes of thermonuclear burning in XRBs may influence the determination of neutron star masses and radii from 
the observation of PRE bursts: via the opacities of the expanding layer (affecting the value of the assumed Eddington limit $L_{Edd}$) and via the interpretation of the observed spectrum (through the deviations of the actual atmosphere spectrum from a perfect blackbody).  Note that other quantities characteristic of neutron stars may also be affected by the composition of burst ashes.  These include thermal \cite{Mir90,Sch99}, radiative \cite{Pac83}, electrical \cite{Bro98,Sch99} and mechanical \cite{BC95,Bil98} properties of the neutron
star crust, which are important for understanding, e.g., 
the evolution of the neutron star magnetic field or the persistent emission between bursts.

\subsection{Burst oscillations: burning in confined regions?}

Oscillations in the light curves of X-ray bursts, with frequencies between about 250 and 600 Hz \cite{sb06}, were first reported by Strohmayer et al. (1996) \cite{stro}
and have now been identified in about 25\% of all bursting sources.  In bursts from LMXBs, oscillations during the rise time may be consistent with the spreading of a hot spot on a rotating neutron star \cite{Str97}.  Oscillations during the burst tails, however, exhibit a drift in frequency, increasing by a few Hz and approaching an asymptotic value as the burst progresses.  The asymptotic frequency of a burst seems to be a characteristic attribute of an individual source, being stable from burst to burst \cite{mun02}.  Oscillations have also been observed in bursts from accreting millisecond pulsars \cite{Wij98}, at frequencies corresponding to the known spin frequencies and with no noticeable drift during the burst tails \cite{cha, str03}.  This suggests that a confined radiating region would be the most straightforward way to account for the oscillations, although more sophisticated explanations are clearly needed to account for the frequency drifts in bursts from LMXBs.  Indeed, altogether different underlying mechanisms may be responsible for the burst oscillations in the two scenarios.  Suggested mechanisms include surface wave modes in the neutron star ocean and atmosphere \cite{heyl, piro, lee, nar}, Coriolis force containment \cite{spit}, and magnetic confinement of the fuel \cite{cav11}.  While the magnetic field in most LMXBs is likely too small ($\sim$10$^8$ G) to confine the fuel, the magnetic field could be enhanced during bursts \cite{bou10}.

\subsection{Multi-peaked X-ray bursts}
\label{multipeak}

Several bursters have exhibited multi-peaked X-ray bursts without evidence of photospheric expansion, where both the bolometric luminosity and the temperature show multi-peaked profiles. These include double-peaked bursts from 4U 1608-52
\cite{pen}, GX 17+2 \cite{kul02} and 4U 1709-267 \cite{jon04}, and about a dozen double-peaked and two unique triple-peaked bursts from the well-studied source 4U 1636-53 \cite{sza,lew87,van86,bs06a, bs06b, gal08,zhang09}. Examples of double-peaked light curves are shown in Fig.~\ref{lcs}.  The separation of the
two peaks in the bolometric light curve is typically $\approx$3--5 seconds, with the flux decreasing to values beween 20\% and 50\% of the maximum peak flux during this interval \cite{gal08,bs06a,bs06b}. Time-resolved spectral analysis shows that the emitting area increases steadily from the start of the bursts until the beginning of the cooling tail \cite{bs06a,bs06b}. As well, a study of double- and triple-peaked bursts from 4U 1636-53
showed that all multi-peaked bursts occurred during periods of high accretion rate \cite{zhang11}.

Bhattacharyya \& Strohmayer \cite{bs06a,bs06b} suggested that aspects of the observed double-peak features could be explained through the propagation of a burning front ignited either at the neutron star rotational pole or at high latitudes. The propagation towards the equator
would be stalled at low latitudes as it would advance against the flow of matter transferred onto the equatorial plane from the accretion disk.  This could explain the cooling after the first peak.  After crossing the
equator, the burning front would advance rapidly again towards the opposite pole, explaining the second peak (but see Ref. \cite{Wat07}). In addition, ignition at the rotational pole would imply the absence of oscillations in the observed emission, as measured in some cases
\cite{bs06a}. Ignition at high latitudes (but not at the pole) could explain the millisecond oscillations detected in the first peak for 4U 1636-53, while the spread of the burning flame throughout the entire surface would explain the disappearance of the
oscillations during the second peak \cite{bs06b}. Other explanations include two-step energy generation due to shear instabilities in the fuel \cite{fuji88} or scattering of the X-ray emission by material evaporated from
the disk \cite{melia92}, but these models do not reproduce the observed double-peaked burst temperature and radius profiles \cite{bs06a}. A nuclear waiting point impedance in the thermonuclear reaction flow (see Section \ref{modresults}) was
suggested by Fisker et al.\cite{Fis04}, but this has difficulties in explaining the large dips observed between the two peaks \cite{bs06a}.

\subsection{Absorption features in X-ray bursts}
\label{absorption}

The detection of absorption lines in the neutron star atmosphere is potentially the most direct and powerful tool for probing both the neutron star equation of state (via determination of the gravitational redshift $z$), and the products of nucleosynthesis
during XRBs (via the determination of abundances in the neutron star atmosphere, especially for PRE bursts \cite{wei06}). The latter consideration is particularly relevant given that no material has been observed to be ejected during XRBs.  The main observational difficulty with the detection of absorption features in XRBs is the short duration of the bursts: long exposure times are required to
accumulate data with enough signal to noise. Even for the most regular bursters such as GS 1826-24, KS 1731-260 and IM 0836-425, with bursts of 20--40 s in duration occurring every 2--3 hours, only $\approx$0.3\% of the total observation time is spent measuring the emission from the bursts.  As well, line broadening effects can exacerbate the difficulties in identifying and interpreting absorption features \cite{Bil03,Vil04,Cha06} (but see also Ref.\cite{intzand10}).

Cottam et al.\cite{cottam2002} reported the first detection of gravitationally-redshifted absorption lines, in the X-ray spectrum of EXO 0748-676.  These results were serendipitously obtained as EXO 0748-676 had been
chosen as a calibration source for the XMM-Newton observatory.  As a result, it was observed for 335 ks with the onboard Reflection Grating Spectrometer; such long periods for spectroscopic observations are not normally possible with X-ray observatories. A total
of 28 type I bursts occurred during this time, with an effective exposure time to the X-ray burst spectrum of only 3200 s. This allowed the authors to identify several absorption features, including lines from Fe XXV, Fe XXVI and O VIII that had
been gravitationally redshifted by $z=0.35$. However, these measurements were not confirmed during a second, longer observation of the same source with the same instrument \cite{cottam_failed}. In addition, detailed
calculations of the expected emission spectrum from the neutron star atmosphere of EXO 0748-676 \cite{Rauch} have cast serious doubts on the ability to identify absorption lines.  A long observation of the regular bursting source GS 1826-24 also
failed to detect any absorption features \cite{kong}.  

Exciting results have recently been reported using RXTE that indicate the observation, albeit with limited spectral resolution, of absorption edges from Fe-peak elements in photospheric radius ``superexpansion" bursts \cite{intzand10}.  Such efforts are highly encouraging and future studies using instruments with higher spectral resolution (such as XMM-Newton or Chandra\cite{CHANDRA}) should be a priority.

\section{Modeling type I X-ray bursts}
\subsection{Defining the mechanism}
\label{mech}

Within a year of the serendipitous discovery of XRBs \cite{gri76,Bel76} more than 20 bursting sources were discovered. This included the enigmatic Rapid Burster
\cite{Lew76,Ulm77}, clearly at odds with the general behaviour exhibited by all the other bursting sources (see Section \ref{intro}). Any model attempting to characterize XRBs had to account simultaneously for the intermediate to long
recurrence periods inferred from the majority of the sources as well as for the extremely short recurrence periods reported for the Rapid Burster.

A first series of models, in which bursts were driven by instabilities in the accretion flow onto a compact, stellar corpse (a white dwarf, a neutron star, or a black hole) was proposed by several groups.  Clearly, a compact object was
required to host the explosion to guarantee that large amounts of gravitational energy were released as X-rays by matter falling into its deep gravitational well.  The exact nature of the underlying compact object was a matter of debate, and in fact one of the early models assumed accretion onto a supermassive black hole ($> 100$ $M_\odot$ \cite{GG76}). A breakthrough was achieved through observations of the bursting source 4U 1724-30 with the OSO-8 satellite \cite{Swa77}: the spectral evolution of
one of its long bursts, best fitted with a blackbody spectrum with kT $\approx$ 0.87 -- 2.3 keV, suggested a much smaller source (i.e., either a neutron star or a stellar black hole). Assuming a source distance of $\approx$ 10 kpc,
Refs. \cite{Hof77a,Hof77b} inferred a blackbody radius for the source of $\approx$ 10 km. Other features such as spectral softening during the decay of type I XRBs, the harder X-ray spectra of bursting sources as compared with most of the X-ray
transients hosting black hole candidates, as well as the masses inferred from those same systems, pointed towards neutron star primaries \cite{vPM95}.  These pioneering simulations considered effects such as instabilities in the interaction between the accreted matter and the neutron star magnetosphere \cite{Hen76,Sve76,Baa77,JR77,Lam77,AC79,Hay81,Hor81}, flare-like eruptions of magnetic energy generated and
released through a toroidal field in the accretion disk \cite{Whe77}, convective-driven instabilities in the disk \cite{Lia77a,Lia77b}, or thermal instabilities driven by Compton heating of the accretion flow \cite{Gri78}. 

A second series of calculations relied on thermonuclear explosions in the layers 
accreted onto a neutron star \cite{WT76,MC77,Jos77,LL78,TP78}.  It was found that the characteristic timescale of a thermonuclear
runaway on top of a neutron star was $\leq$ 1 s\cite{HvH75}, in rough agreement with the timescale of the X-ray emission reported from some bursting sources.  It was clear, however, that 
the quick succession of flashes observed from the Rapid Burster could not be accounted for by this thermonuclear 
runaway model since not enough fuel could be piled up on the surface of the 
neutron star in such short time intervals.  Progress was made through 
the subsequent observation of two different types of bursts associated with the Rapid Burster, with different recurrence periods \cite{Hof77c,Hof78}. 
The classification of type I and type II bursts (the former associated with 
thermonuclear flashes, the latter to accretion instabilities) was then suggested.

\subsection{The thermonuclear runaway model}

Rosenbluth et al.\cite{Ros73} were the first to quantify the nuclear energy released
from the fusion of H-rich material on accreting neutron stars. It was later concluded \cite{vHH74,HvH75} that under certain conditions such nuclear fusion episodes may trigger
thermonuclear runaways. The connection between unstable nuclear burning on the
surfaces of neutron stars and type I XRBs, as envisaged by Woosley and Taam \cite{WT76}
 (for He-driven bursts) and Maraschi and Cavaliere \cite{MC77} (for H-driven
bursts)\footnote{According to Ref. \cite{Lew95}, the earliest suggestion of
  the thermonuclear origin of type I XRBs was made by L. Maraschi while visiting MIT in February 1976, at the time when the central 
  region of the Milky Way was being observed with the SAS-3 satellite\cite{SAS}.} paved 
the way for a suite of increasingly detailed numerical simulations \cite{Jos78,
TP79, CJ80,JL80,ET80,Taa80,Taa81,Fuj81,Taa82,AJ82,Pac83,WW84}. 
Initial efforts to compare theoretical predictions and observations were made by Refs. \cite{Jos77,LL78}. 

The gross observational features of type I XRBs were qualitatively reproduced
by early studies, already suggesting the possibility of constraining neutron star properties (e.g., masses, radii, equation of state) through models.  Early dimensional 
analysis work \cite{Jos77,LL78} revealed that for bursts driven by He-burning, one could expect recurrence periods of about 10 000 s, 
accreted envelope masses of the order of $\approx 10^{21}$ g, and an overall
energy release of $\leq 10^{39}$ erg per burst.  Later predictions included peak luminosities of $\approx 10^{38}$ erg 
s$^{-1}$ (with peak blackbody temperatures of $3 \times 10^7$ K),
light curve rise times of $\approx 0.1$ s, and light curve decay times of $\approx$ 10 s, through consideration of the transport of nuclear 
energy from deep in the envelope to the outermost neutron star 
layers.
Another important observational constraint matched by thermonuclear models
of type I XRBs was the $\alpha$ parameter (see Section \ref{modresults}), the ratio between time-integrated
persistent and burst fluxes. 

The first detailed numerical models of XRBs can be found in Joss~(1978) \cite{Jos78} for 
several sets of mass-accretion rates and neutron star central temperatures.  (The central temperature determines the initial 
  surface luminosity assumed in a model.)
These parameters, together with the neutron star mass, equation of state and the 
metallicity of the accreted material, represent key ingredients in XRB modeling.
Joss's work quantitatively confirmed the values derived from dimensional 
analysis.  As well, it stressed that a burst consumes virtually all the accreted fuel (likely 
synthesizing Fe-peak elements) and that the energy released during the 
explosion is essentially emitted as X-rays from the neutron star photosphere.  Further numerical simulations \cite{JL80}
revealed that for highly magnetized neutron stars ($B \geq 10^{12}$ G), matter 
transferred from the stellar companion would be funneled onto the neutron star 
magnetic poles, enhancing the local accretion rates in those regions by a 
factor of $\approx 10 ^3$.  Ignition regimes as a function of mass-accretion rate were later defined \cite{Fuj81,FL87} for the accretion of material of solar metallicity (see Section \ref{obsover}).

General relativistic corrections were first incorporated into XRB models by Refs. \cite{TP79,CJ80,Taa80}, and more consistently by 
Ref. \cite{AJ82}. These are important for the interpretation of observational properties of X-ray bursts. For example, previously predicted peak luminosities and burst recurrence times were reduced and increased by a factor of $1+z$, respectively, where $z$ is
the gravitational redshift (see Section \ref{obsover}).

\subsection{Model predictions of nucleosynthesis in XRBs}
\label{modpred}
\subsubsection{Overview}
\label{modover}

Before discussing results from XRB models, one must carefully distinguish between the different numerical 
approaches used in the calculations.  One category of models includes
parameterized one-zone calculations \cite{Fuj81,WW81,Han83,Has83,Koi99,Sch99,Koi04, 
Sch01}. These simple prescriptions relate the history of the neutron star's accreted
envelope with the time evolution of the temperature $T$ and density $\rho$ in a 
single layer of the envelope.  Such 
thermodynamic quantities are often determined through semi-analytical 
models, or occasionally correspond to T-$\rho$ profiles directly extracted from
1-D hydrodynamic models.  This approach, while representing an extreme 
oversimplification of the physical conditions governing neutron star envelopes,
was used extensively to overcome the time limitations that arose in 
computationally intensive hydrodynamic calculations.  For similar reasons, one-zone calculations 
have been used recently to estimate the impact of nuclear 
uncertainties on the final XRB yields \cite{Amt06,Rob06,Par08,Par09} (see Section \ref{sens}). 

At the moment, state-of-the-art XRB nucleosynthesis calculations rely on 1-D 
hydrodynamic models (see Refs. \cite{Wal82,WW84,Taa93,Taa96,Woo04,Fis08,Jos10}, and 
references therein).  The underlying assumption in all these models is spherical
symmetry. This simplifying hypothesis demands that the explosion occur
uniformly over a spherical shell.  Moreover, spherically symmetric models are limited by their treatment of the manner in which thermonuclear runaways are
initiated (presumably as point-like ignitions) and propagate. This particularly 
affects convective mixing, a key ingredient in deflagrative, explosive 
scenarios such as type I X-ray bursts that can only be accurately modeled in
three dimensions.  Multidimensional XRB models are computationally challenging 
and only preliminary simulations of specific aspects of the explosions (such as flame propagation\cite{spit} or the early convective stages preceding 
ignition\cite{Mal11}) have been conducted to date. Hence, no reliable 
nucleosynthetic predictions have been reported using multidimensional simulations.

The relevant nuclear reaction sequences accompanying XRBs were initially identified in the 1980s--90s \cite{WW81,vW94,Sch99,Ili99}. Because of 
computational constraints, early attempts to predict the nucleosynthesis 
expected during XRBs were performed using limited nuclear reaction networks, 
truncated near Ni (and using a 19-isotope network) \cite{WW84,Taa93,Taa96}; Cu (181-isotope network) \cite{Has83}; Kr (274-isotope \cite{Han83} or 463-isotope networks \cite{Koi99}); Cd (16-isotope network) \cite{Wal84}; or Y 
(250-isotope network) \cite{WW81}. Efforts to link state-of-the-art hydrodynamic models
with both accurate input physics and large nuclear reaction networks were clearly 
needed but required advanced generations of computers not available until recently.

\subsubsection{Results from models with large nuclear reaction networks}
\label{modresults}

With a neutron star as the underlying compact object hosting a thermonuclear explosion, 
temperatures and densities in the accreted envelope reach $T > 10^9$ K 
and $\rho \approx 10^6$ g cm$^{-3}$.  Detailed nucleosynthesis 
studies under such conditions require the use of networks with hundreds of isotopes and thousands of nuclear 
interactions.  Pioneering efforts in this
regard were made by Schatz et al. \cite{Sch99,Sch01}, who carried out XRB 
nucleosynthesis calculations with a network containing more than 600 isotopes 
(up to Xe).  Unfortunately, those calculations relied on
a one-zone approach and so inadequate attention was given to the stellar model (and to convection in particular). 
Koike et al.\cite{Koi04} also performed detailed one-zone nucleosynthesis 
calculations, using temperature and density profiles extracted from spherically 
symmetric hydrodynamic models.  These profiles were linked to a 1270-isotope network extending up to $^{198}$Bi.  Despite the limitations in the adopted 
approach, these works confirmed that the main nuclear reaction flow is driven by
the rp-process (rapid proton-captures and $\beta^+$-decays), the 
3$\alpha$-reaction, and the $\alpha$p-process (a sequence of ($\alpha$,p) and 
(p,$\gamma$) reactions).  They revealed that the abundance flow proceeds far away from the 
valley of stability, merging with the proton drip-line beyond A = 38.  They also
stressed the need to rely on huge networks, since the main nuclear path extends
up to the SnSbTe mass region \cite{Sch01} or even beyond (e.g., the 
nuclear activity in Ref. \cite{Koi04} reached $^{126}$Xe). It is clear that the location of the nucleosynthetic endpoint in XRB simulations depends upon the astrophysical parameters adopted.  The specific 
location of the maximum possible endpoint in the most extreme bursts is still a matter 
of debate, since recent experimental studies \cite{Elo09a} have revealed
difficulties in reaching the SnSbTe-mass region.

Only very recently has it been possible to couple hydrodynamic models and
detailed nuclear reaction networks.  Efforts include Refs. \cite{Fis04,Fis06,Fis07,Fis08,Tan07}, with networks of $\approx$300 isotopes (up to Te); Refs.
\cite{Woo04,Heg07}, with networks of 1300 isotopes and an 
adaptive network; and Refs. \cite{JM06,Jos10}, with networks of 325 isotopes (up to Te).  The first set of calculations \cite{Fis04,Fis06,Fis07,Fis08,Tan07} relied upon the 
1-D, fully relativistic hydrodynamic code AGILE, with 129 numerical shells and convective 
transport based on mixing-length theory.  In turn, the second set \cite{Woo04,Heg07} was performed with the 1-D, Newtonian hydrodynamic KEPLER code, with a variable number of
shells (up to 1000 for some models), time-dependent convective transport based
on mixing length theory with overshooting, and thermohaline mixing.  Finally, the third set of calculations \cite{JM06,Jos10} used the 1-D, Newtonian hydrodynamic code SHIVA, with up to 200 numerical shells
and convective transport based on time-dependent mixing-length theory.  Overall, a number of XRB observables have been reproduced by current hydrodynamic models. This includes recurrence periods, peak luminosities and $\alpha$ values (see, e.g., Heger et al. (2007) \cite{heg07} for a comparison between model predictions and the regular bursts from GS 1826-24).  Nonetheless, certain aspects associated with the observed shapes of the light curves, such as the rise times, the presence of bolometric light curves with multiple peaks, or the incredible diversity in light curve shapes observed from individual sources, have not been fully understood yet.

Here we summarize results obtained by the Barcelona group \cite{Jos10} using models of 1.4 M$_\odot$ neutron stars accreting matter of different metallicities at a rate of $1.75 
\times 10^{-9}$ M$_\odot$
yr$^{-1}$ (0.08 $\dot M_{Edd}$). The initial luminosity of the neutron star, at the onset of accretion, is $L_{initial} = 4.14$ L$_\odot$.  We outline the most striking features of the main nuclear activity during X-ray bursts and compare these results with those from simulations performed by the Santa Cruz\cite{Woo04} and Basel\cite{Fis08} groups.  Interested readers can find more detailed accounts of the dominant flows at different stages of the bursts in Refs.\cite{Woo04,Fis08,Jos10}.  

For models accreting material of solar metallicity, results include burst recurrence times of $\tau_{rec} \approx 5$ hr, peak 
temperatures $T_{peak}$ of 
$\approx 1.1-1.3$ GK, peak luminosities $L_{peak} = (4-8) \times 
10^{38}$ erg s$^{-1}$ and ratios between persistent and burst 
luminosities in the range of $\alpha = 30-40$.
While these values are qualitatively in agreement with those inferred from the 
bursting sources GS 
1826-24 ($\tau_{rec}= 5.74 \pm 0.13$ hr; $\alpha = 41.7 \pm 1.6$), 4U 1323-62 
($\tau_{rec}= 5.3$ hr; $\alpha = 38 \pm 4$), or  4U 1608-52
($\tau_{rec}= 4.14 - 7.5 $ hr; $\alpha = 41 - 54$),
the simulations systematically yield larger recurrence times 
 and peak luminosities (and hence, 
lower $\alpha$) than those reported by the Santa Cruz group \cite{Woo04} for their model ZM.  The twelve
bursts computed in Ref. \cite{Woo04} are characterized by recurrence times of about 
$\approx 2.7$ hr, peak luminosities of $L_{peak} \approx (1.5 - 2) \times 10^{38}$ 
erg s$^{-1}$, and ratios between persistent and burst luminosities of  
$\alpha \approx 60-65$.  We note that models from the Basel group \cite{Fis08} give lower $L_{peak}$ and somewhat larger $\alpha$ values than models from both the Barcelona and Santa Cruz groups.  
  
\begin{figure}  
\begin{center}  
\epsfxsize=6in {\epsfbox{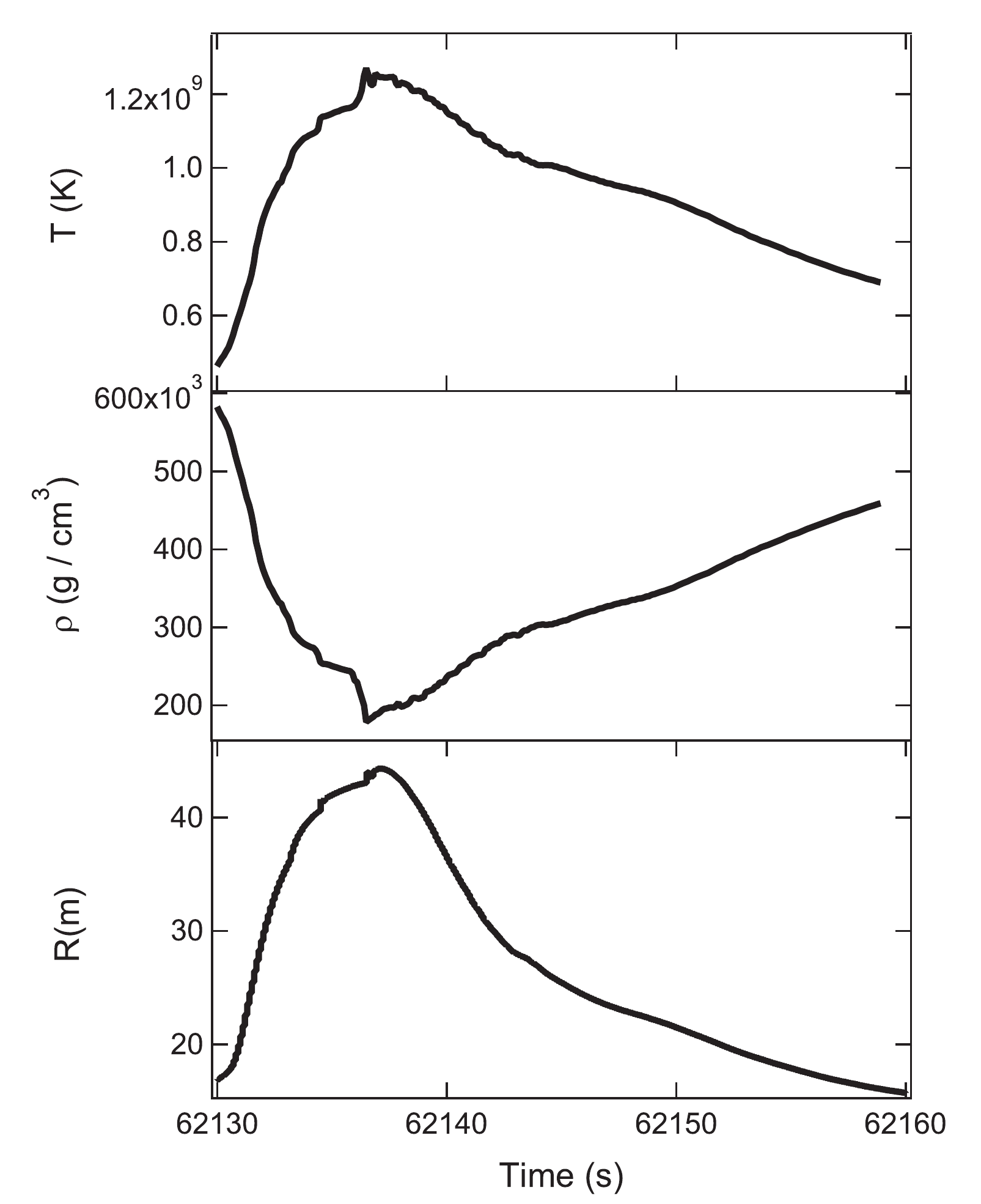}}  
\caption{(Top) Temperature and (middle) density versus time during the evolution of the hottest shell from a hydrodynamic XRB model (burst 3 of model 1 from Ref. \cite{Jos10}).  (Bottom) Envelope size during the same burst, measured relative to the core-envelope interface.  Note the modest expansion of the envelope during the burst, achieving a maximum height of 44 m above the core-envelope interface.  The associated light curve for this burst is shown in Fig.~\ref{comp_light_curves}.}
\label{profiles}  
\end{center}  
\end{figure}

\begin{figure}  
\begin{center}  
\includegraphics[scale=1.0]{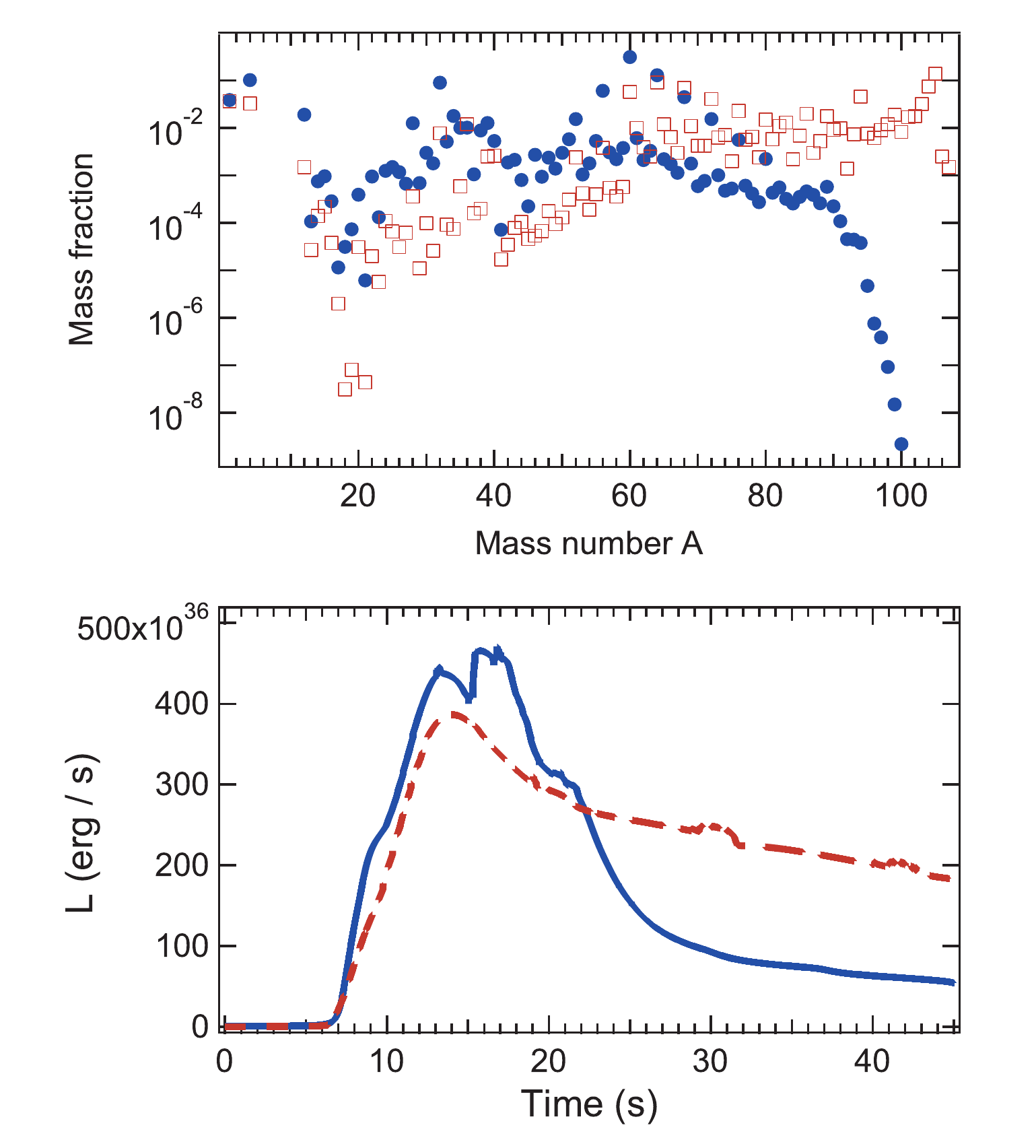}
\caption{Effect of the metallicity of the accreted material
on type I X-ray bursts.  (Top, filled circles) Mean composition of the envelope following the fourth burst computed in a model 
with a 1.4 M$_\odot$ neutron star, accreting solar-like material (Z=0.02) at 
a rate of $1.75 \times 10^{-9}$ M$_\odot$ yr$^{-1}$ \cite{Jos10};  (bottom, solid line) the light curve associated with this burst.  (Top, open squares) Mean composition of the envelope following the fifth burst computed in a similar model as above, but accreting material of lower metallicity (Z=$10^{-3}$)\cite{Jos10}; (bottom, dashed line) the light curve associated with this burst.  Accreted material of lower metal content drives bursts characterized by lower 
peak luminosities and longer decline times. In turn, the longer duration of the burst
allows a significant extension of the nuclear activity towards the SnSbTe-mass region,
which is clearly reflected in the mean composition of the envelope at the end of the
burst.}
\label{Ltonuc}  
\end{center}  
\end{figure}

The time evolution of temperature and density at the ignition shell
during the third bursting episode of the model from the Barcelona group is shown in Fig.~\ref{profiles}, top and middle panels. Note the modest
expansion of the envelope during a burst (Fig.~\ref{profiles}, lower panel), which takes
place at nearly constant pressure.  The moderate $T_{peak}$ achieved in 
this model restricts most of the nuclear activity to $A \approx 60$ (mainly 
$^{60}$Ni and $^{64}$Zn \cite{Jos10}) and no 
large concentrations of material in the SnSbTe-mass region were found, in agreement
with the results obtained by the Santa Cruz group \cite{Woo04}.  The mean composition of the envelope at the end of the fourth burst in this model is illustrated in the top panel of Fig.~\ref{Ltonuc} (filled circles); the corresponding light curve is given in the lower panel of Fig.~\ref{Ltonuc} (black line).    
 This model also yields a very small post-burst abundance of $^{12}$C, 
below the threshold amount required to power superbursts 
($X(^{12}C)_{min} >$ 0.1, see Ref. \cite{Cum01}).

Results from a second set of models, identical to those described above but considering
accretion of metal-defficient matter
(Z = Z$_\odot$/20 = 0.001), include longer recurrence times of $\tau_{rec} \approx 9$ hr, $T_{peak}$ of about $1.3-1.4$ GK, $L_{peak} \approx 10^{38}$ erg s$^{-1}$ and ratios between persistent and burst luminosities of $\alpha \approx 20-30$.  These values are comparable to those
measured in the XRB sources 1A 1905+00 ($\tau_{rec}$ = 8.9 hr), 4U 1254-69 
($\tau_{rec}$ = 9.2 hr), or XTE J1710-281 ($\tau_{rec}$ = 8.9 hr, $\alpha = 22-100
$).  The longer exposure to high temperatures (driven by a 
slower decline phase) causes an extension of the main nuclear path towards 
the SnSbTe-mass region, with final abundances at the end of the fifth burst
dominated by the presence of $^{105}$Ag, $^{64}$Zn, $^{104}$Pd and $^{68}$Ge.  The mean composition of the envelope at the end of the fifth burst in this model is illustrated in the top panel of Fig.~\ref{Ltonuc} (open squares); the corresponding light curve is given in the lower panel of Fig.~\ref{Ltonuc} (grey line).       
Similar to the solar metallicity model, little $^{12}$C remains at the end of a burst.

\begin{figure}  
\begin{center}  
\includegraphics[scale=0.5]{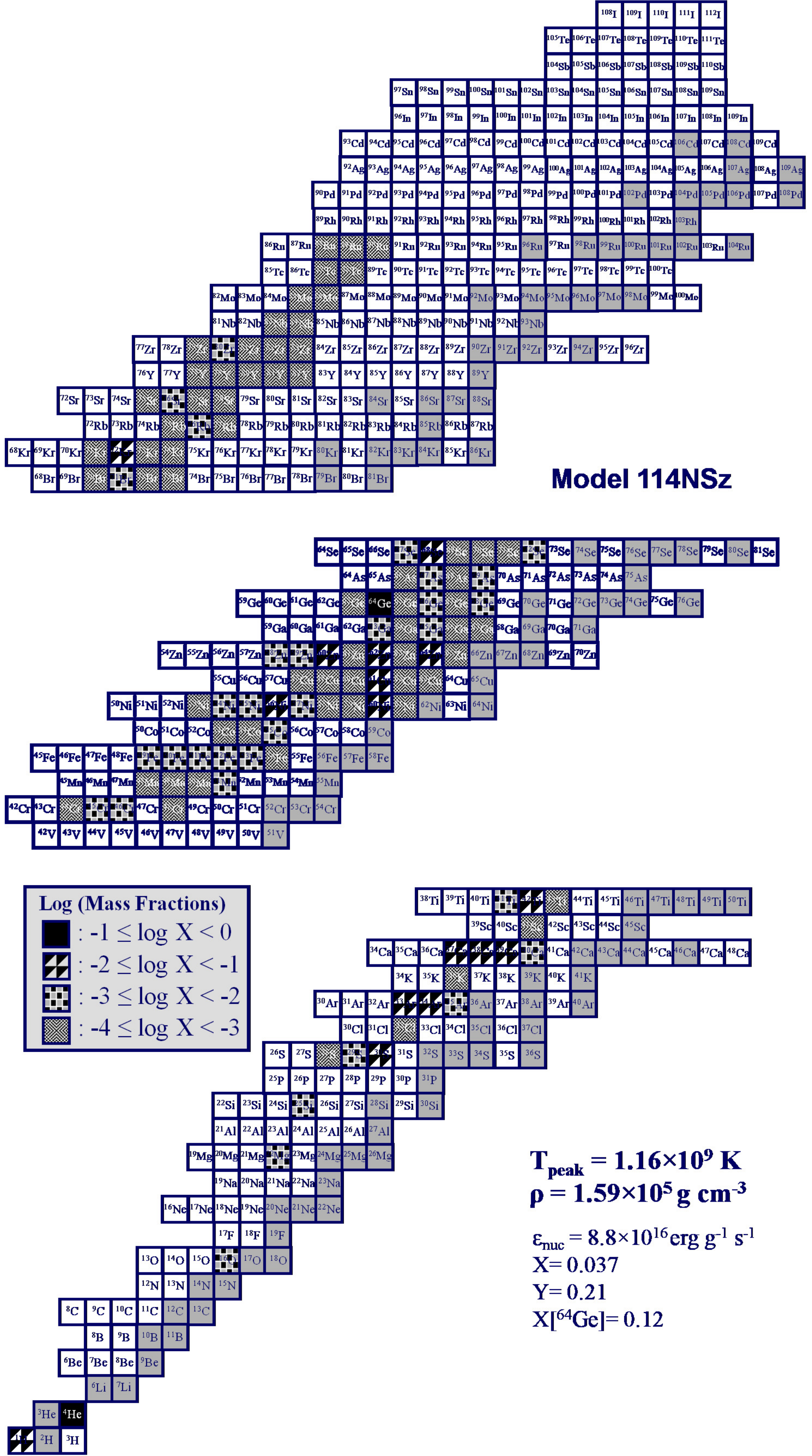}
\caption{Main nuclear activity during a XRB at $T_{peak}$, shown in terms of mass fractions for the most
abundant species ($X \geq 10^{-4}$) within the ignition shell 
($\approx 5.6$ m above the core--envelope interface).   These results were found for the fourth burst computed in a model 
with a 1.4 M$_\odot$ neutron star, accreting solar-like material (Z=0.02) at 
a rate of $1.75 \times 10^{-9}$ M$_\odot$ yr$^{-1}$ \cite{Jos10}. 
Note the large amount of $^{64}$Ge synthesized and the modest extension of the activity 
up to A = 90.}
\label{xzfull}  
\end{center}  
\end{figure}

\begin{figure}  
\begin{center}  
\includegraphics[scale=0.5]{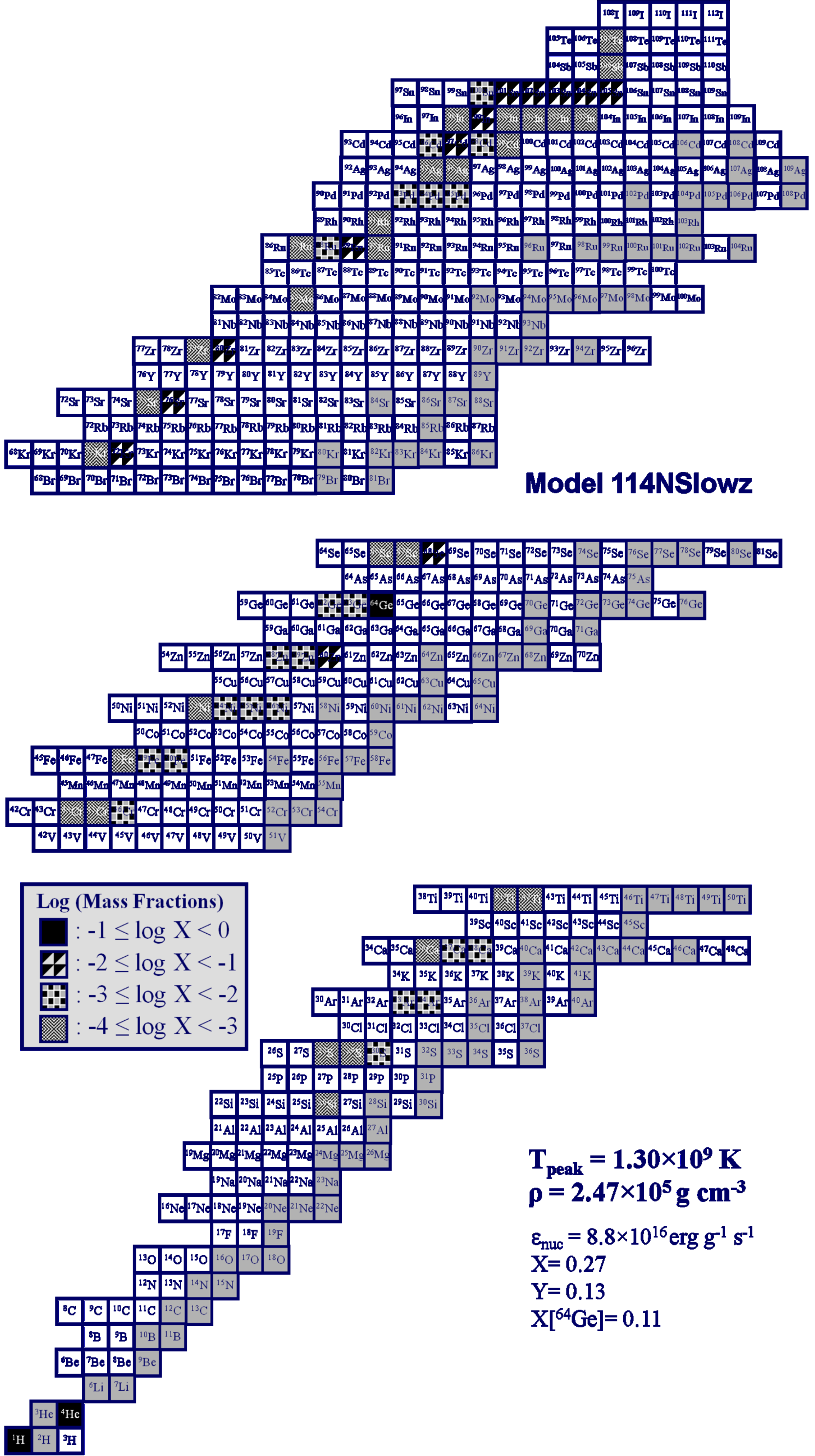}
\caption{Same as Fig.~\ref{xzfull}, but for accretion of low-metallicity (Z=$10^{-3}$) material.
As in Fig.~\ref{xzfull}, the most abundant species
at $T_{peak}$ is $^{64}$Ge; however, the nuclear activity here extends up to
A = 107.}
\label{xlowzfull}  
\end{center}  
\end{figure}

\begin{figure}  
\begin{center}  
\includegraphics[scale=0.5]{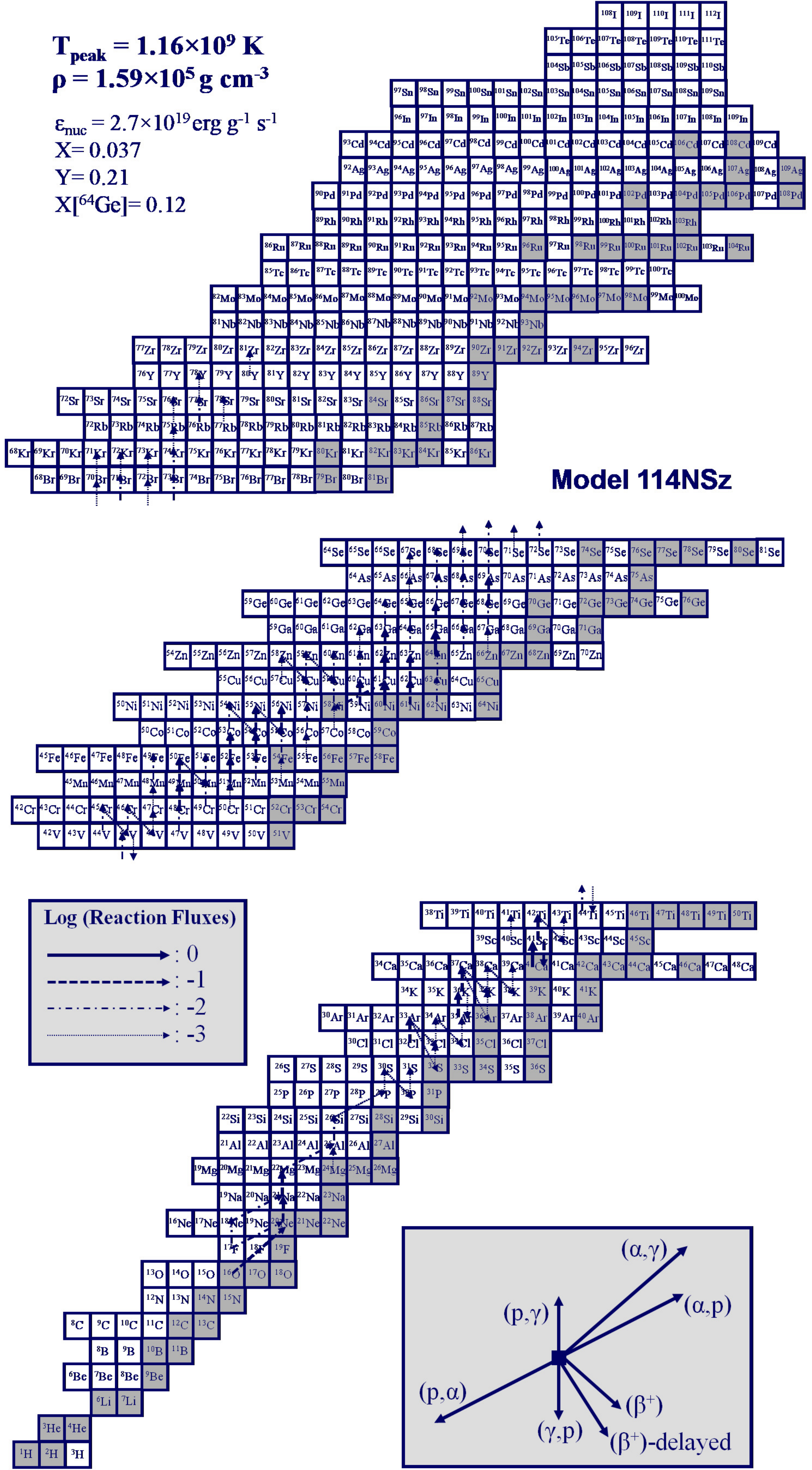}
\caption{Same as Fig.~\ref{xzfull}, but the nuclear activity is shown in terms of the dominant reaction fluxes ($F > 10^{-3}$). For clarity,
the numerous (p, $\gamma$)-($\gamma$,p) pairs at equilibrium are not displayed. Note
that at $T_{peak}$, the nuclear flow is dominated by proton-capture reactions (exceptions
being $^{16}$O($\alpha$,$\gamma$),  $^{17}$F($\alpha$,p), $^{18}$Ne($\alpha$,p), 
and $^{22}$Mg($\alpha$, p)). A few photodisintegration reactions (not in equilibrium with
the corresponding (p,$\gamma$) processes), and a number of $\beta$ and $\beta$-delayed
decays are also shown. The main nuclear activity proceeds
far from the valley of stability above A = 45.}
\label{fzfull}  
\end{center}  
\end{figure}

\begin{figure}  
\begin{center}  
\includegraphics[scale=0.5]{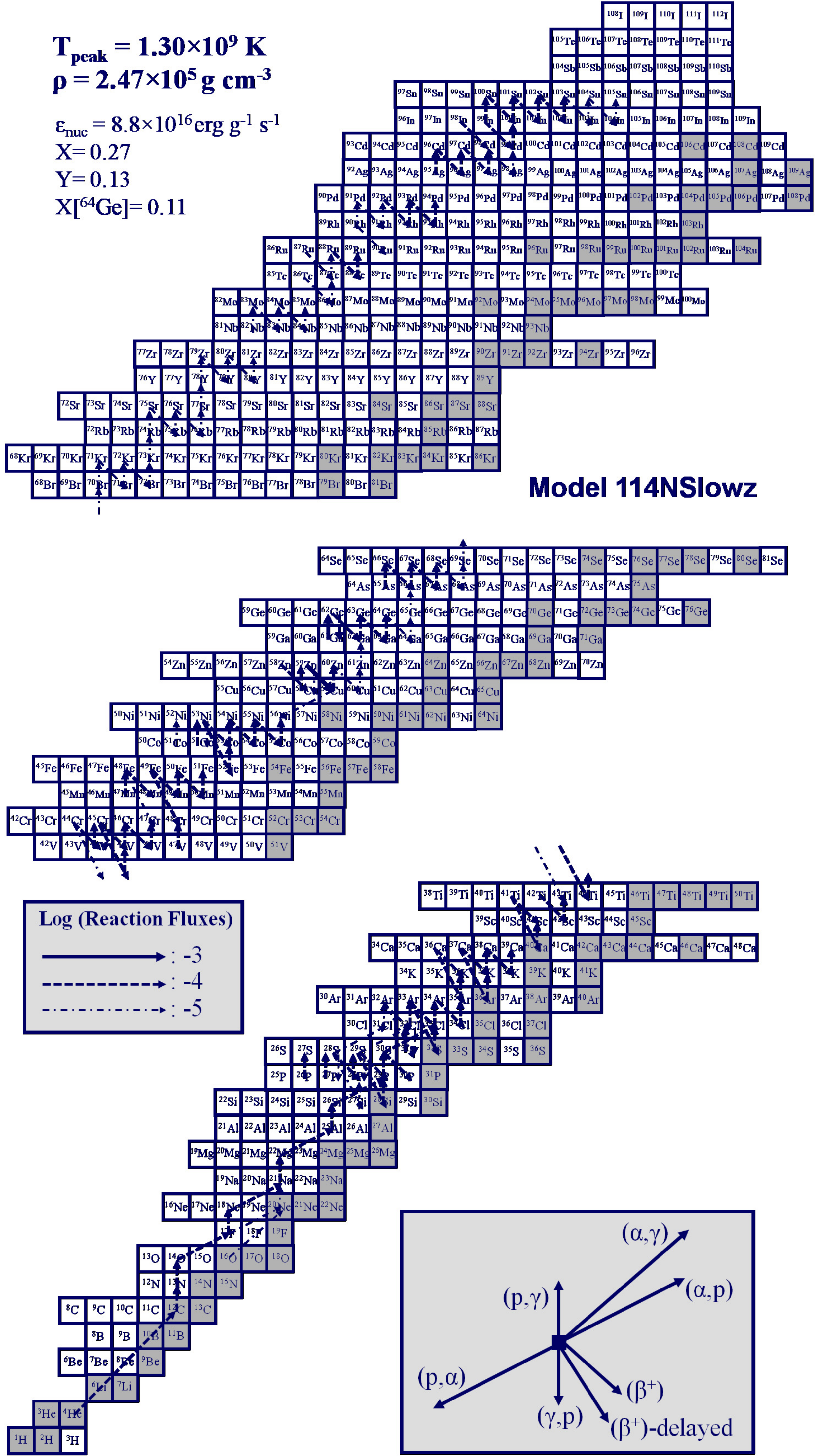}
\caption{Same as Fig.~\ref{xlowzfull}, but the nuclear activity is shown in terms of the dominant reaction fluxes. For clarity,
the numerous (p,$\gamma$)-($\gamma$,p) pairs at equilibrium are not displayed.
The reduced metallicity of the accreted matter results in a net reduction of
the reaction fluxes, with the most abundant fluxes being three orders of magnitude less 
frequent than those reported for the solar metallicity model in Fig. \ref{fzfull}. Note the significant number of $\beta$-decays presented here (compare Fig.~\ref{fzfull}) and the extension of
the activity up to the SnSbTe-mass region.}
\label{flowzfull}  
\end{center}  
\end{figure}

The two models from the Barcelona group discussed above differ only in the metallicity of the accreted material.  The nuclear activity in these two models is compared in detail in Figs.~\ref{xzfull}, \ref{xlowzfull}, \ref{fzfull} and \ref{flowzfull}.  These plots show the most abundant species ($X>10^{-4}$) and dominant reaction fluxes at $T_{peak}$, within the ignition shell of the fourth computed burst in each model.
Figs.~\ref{xzfull} and \ref{xlowzfull} reveal the existence of {\it waiting point nuclei}: unstable nuclei (e.g., $^{30}$S, $^{56}$Ni, $^{60}$Zn, $^{64}$Ge, $^{68}$Se, $^{72}$Kr, $^{76}$Sr, $^{80}$Zr, $^{89}$Ru, etc.) where the abundance flow is temporarily delayed, allowing material to accumulate.  The manner by which each waiting point nuclide is eventually depleted depends on the nuclear structure in the immediate mass region of this species:
some are destroyed by $\beta$-decay, some by ($\alpha$,p) reactions, and others by sequential (rather than direct) two-proton capture (see Section \ref{nucover}).  The bursts in both models are eventually quenched by the exhaustion of H and He fuel rather than through any significant envelope expansion (see Fig.~\ref{profiles}, lower panel).  We note that a typical range of metallicity for the accreted material is difficult to determine. Whereas solar metallicities may be regarded as representative (e.g., a thorough comparison between observations of the X-ray burster GS 1826-24 and models with different metallicities favoured the accretion of material of solar metallicity\cite{heg07}), the identification of many bursting sources in globular clusters implies that  lower metallicities should also be expected.

Comparison of results from the solar and low-metallicity models suggests that the smaller the metal content of the accreted material, the larger the 
recurrence time (and the smaller the $\alpha$).  In addition, explosions in 
metal-deficient envelopes are characterized by lower peak luminosities and 
larger light curve decline times.  Similar trends are reported by the Santa Cruz group \cite{Woo04} when comparing models accreting solar and low metallicity material.  This is in spite of any systematic differences between the models, as mentioned above. The most striking difference between the models of Ref. \cite{Woo04} and Ref. \cite{Jos10} concerns the larger effect 
played by the metallicity of the accreted material in the models computed
by the Barcelona group \cite{Jos10}.  Indeed, Ref. \cite{Woo04} 
 explained the moderate effect they found as due to ``compositional inertia" washing out the influence of the initial metallicity.  This effect refers to how properties of bursts may be sensitive to the fact that
accretion occurs onto the ashes of previous bursts, minimizing the impact of the actual metallicity of the accreted material after sufficient bursts have been calculated.  Similar conclusions
were reached by Ref. \cite{Fis08}.
The origin of these discrepancies (as well as other discrepancies in peak luminosities, recurrence rates or $\alpha$ values for a common metallicity) is not yet clear. Although XRB 
properties depend only weakly upon the neutron star mass (or surface gravity), some of the differences outlined between the three studies may be 
attributed to the combined effects of different assumptions for the adopted neutron star
size and to differences in the input physics (e.g., nuclear reaction network, opacities, treatment of convective transport). 

Aside from any metallicity-dependent effects, it is evident from Fig.~\ref{Ltonuc} that different nucleosynthesis patterns are associated with different characteristic burst light curves.  This is seen even within sequences of bursts calculated at a given metallicity (see, e.g., Figs. 19 and 20 in Ref. \cite{Jos10}, or Figs. 13 and 15 in Ref. \cite{Woo04}).  This relationship is of vital importance since the products of nucleosynthesis during an XRB have not been directly observed in a straightforward, reliable manner (see Section \ref{absorption}).  As such, from the point of view of nuclear astrophysics, experimental nuclear physicists cannot rely solely on exploring the effect of a nuclear laboratory measurement on XRB nucleosynthesis; one must also examine the effect of the measurement on the principal observable of an XRB, namely, the burst light curve.  This is in stark contrast to the cases of classical nova explosions or supernovae, in which the products of nucleosynthesis are ejected into space and can be observed through spectroscopy or other means.  Fortunately, the link between nucleosynthesis in XRBs and the predicted light curves provides an essential guide to help judge which nuclear physics uncertainties may most significantly affect XRB light curves (see Section \ref{sens}).       

The effect of the adopted mass-accretion rate on burst properties has recently 
been investigated by Ref. \cite{Woo04}.  Models  
 of 1.4 M$_\odot$ neutron stars ($L_{initial} = 4.14$ L$_\odot$)
accreting matter at $3.5 \times 10^{-10}$ M$_\odot$
yr$^{-1}$ (0.02 $\dot M_{Edd}$) and at $1.75 \times 10^{-9}$ M$_\odot$
yr$^{-1}$ (as mentioned above) were computed for two different
metallicities (solar and 5\% solar).  For the models accreting material of solar metallicity, those with $\dot M = 1.75 \times 10^{-9}$ M$_\odot$ exhibited combined H/He flashes, while bursts from models with 
 $\dot M = 3.5 \times 10^{-10}$ M$_\odot$ yr$^{-1}$ experienced ignition in a pure, H-free, He layer, yielding 
briefer, brighter light curves with shorter tails, very rapid rise times
($<0.1$ s), and ashes corresponding to elements below the Fe region.  This is in agreement with the discussion in Section \ref{obsover}.  

With regard to bursts featuring bolometric luminosities with double-peaked profiles, Fisker et al.\cite{Fis04} report that some of the observed features (see Section \ref{multipeak}) could be explained, in part, through a waiting point impedance in the nucleosynthesis accompanying an XRB.  For suitable accretion rates, the first peak could be caused by the rapid ignition of He accumulated at the base of the envelope, following the burning of H through the CNO cycle.  Energy released from this He burning region would stimulate thermonuclear ignition in other shells where H may not have been fully converted to He.  The abundance flows in these higher-lying shells may then determine whether or not a second peak is produced.  For example, if the flow encounters a waiting point nucleus, the nuclear energy generation rate may decrease, possibly causing the surface luminosity to also decrease.  Once the flow proceeds again, beyond the waiting point, the surface luminosity may increase again.  Fisker et al. suggest that waiting points at $^{22}$Mg, $^{26}$Si, $^{30}$S, and $^{34}$Ar may be of consequence in this context, and that reaction rates to better determine how these nuclei are eventually destroyed (e.g., through ($\alpha$,p) reactions) should therefore be measured.             

Finally, a long-debated aspect of the nucleosynthesis accompanying XRBs is the
potential role of these phenomena as nuclear factories of the elusive light p-nuclei $^{92,94}$Mo 
and $^{96,98}$Ru \cite{Sch98,Sch01,Dau03,wei06,Baz08} (for a review on the production of p-nuclei, see Arnould and Goriely\cite{Arn03}).  Matter accreted onto 
a neutron star of mass $M_{NS}$ $\approx$ 1.4 M$_\odot$ and radius $R_{NS}$ $\approx$ 10 km releases $G M_{NS} m_p/R_{NS} \approx 200$ MeV per
nucleon, whereas only a few MeV per nucleon are released from thermonuclear 
fusion.  Ejection from a neutron star is therefore energetically 
unlikely, as confirmed by all recent hydrodynamic simulations 
\cite{Woo04,Fis08,Jos10}\footnote{The fact that no mass escapes the computational domain, even
in Lagrangian frameworks (i.e., simulations in which the computational grid is 
attached to the fluid), allows one to simulate sequences of 
repeated bursts (see pioneering models by Refs. \cite{Fuj85,WW84}). This has opened up new possibilities to study
the long-term evolution of bursting sources and to identify asymptotic regimes for suitable combinations of mass-accretion rates and
metallicities (see discussion in Ref. \cite{Woo04}).}.
Even though radiation-driven winds during 
photospheric radius expansion (see Section \ref{PRE}) may lead to the ejection of a tiny fraction 
(i.e., $\approx$1\%) of the envelope \cite{Kat83,PP86,Pac90,Nob94,wei06}, it has not yet been rigorously analyzed whether such winds may contain material synthesized during the burst.  Furthermore, in one-zone nucleosynthesis calculations the chemical species are assumed to represent the whole (by 
construction, chemically homogeneous) envelope.  In multi-zone hydrodynamic 
simulations, however, the abundances of many species, including these p-nuclei, decrease dramatically towards the outer envelope layers because of limited 
convective transport.  Unfortunately it is these layers that are most likely to be ejected by radiation-driven winds.  The predicted overproduction factors in these outer regions \cite{Jos10} are several orders of 
magnitude smaller than those required to account for the origin of Galactic 
light p-nuclei \cite{Baz08}, in sharp contrast with the results 
reported from one-zone calculations \cite{Sch01}.  We conclude that XRBs are unlikely to be dominant contributors to the Galactic abundances of p-nuclei, according to current models.

\section{Nuclear physics in type I X-ray bursts}
\subsection{Reaction rate formalism}
\label{nucover}

We have already discussed in Section \ref{modresults} some general features of type I X-ray burst nucleosynthesis. The interplay of charged-particle reactions and $\beta$-decays at temperatures and densities characteristic of XRBs crucially defines the
overall abundance flow and nucleosynthesis. Instead of a smoothly continuous abundance flow from lighter to heavier nuclear species we should envision a flow in very fast spurts ($<$1 s), each delayed significantly at the ``waiting point nuclei" where the capture of successive charged-particles is inhibited by strong reverse photodisintegration
reactions. From these general considerations we can deduce the nuclear quantities generally needed for studying the nucleosynthesis in type I XRBs.

The basic quantity required is the thermonuclear reaction rate per particle pair, $\left<\sigma v\right>$, defined as the number of nuclear reactions taking place in the stellar plasma per unit volume and per unit time.
Specifically, the reaction rate represents the integral over all kinetic energies $E$, of the product of the Maxwell-Boltzmann distribution and the nuclear reaction cross section $\sigma$. It is given by
\begin{equation}
N_A \langle \sigma v \rangle = \left(\frac{8}{\pi m}\right)^{1/2} \frac{N_A}{(kT)^{3/2}}\int_0^\infty E\,\sigma(E)\,e^{-E/kT}\,dE	\label{reactionRate}
\end{equation}
where $m$ is the reduced mass, $N_A$ is Avogadro's number, $k$ is the Boltzmann constant, and $T$ is the stellar temperature. There are a number of strategies for estimating the thermonuclear reaction rate, depending on the nuclear reaction mechanism. When the cross section varies smoothly with energy and is devoid of any structure, the reaction most likely proceeds via a very fast (i.e., less than 10$^{-20}$ seconds), single step mechanism. On the other hand, if the incoming projectile is absorbed by the target nucleus, sharing all its energy with the target nucleons over a much longer period of time (i.e., about 10$^{-16}$ seconds), the cross section most likely will exhibit a sharp peak, called a resonance. In reality many resonances with different energies and strengths (i.e., the integrated resonance cross section) will contribute to the reaction rate. In the most favourable case, the total reaction rate can be derived from experimental nuclear data.  Significant advances have been made recently for evaluating such ``experimental" reaction rates using a Monte Carlo procedure in order to estimate
statistically rigorous values \cite{Ili07,Lon10}. On the other hand, when no nuclear data are available, the reaction rate must be estimated with the aid of nuclear reaction models, for example, the statistical (Hauser-Feshbach) model (see below), or the nuclear shell model (see e.g., \cite{Fis01}).  In either case, reaction rates derived under the assumption that all target nuclei are in their ground states (``laboratory rates") must, in general, be corrected for thermal excitations of the target nuclei in order to estimate ``stellar rates".
The required correction, or ``stellar enhancement factor" (SEF), can also be derived from the Hauser-Feshbach model (see, for example, Refs. \cite{Rau00,Gor08}). For the specific case of type I X-ray bursts, the peak temperatures achieved
($<2$ GK) result in small predicted correction factors for most nuclear reactions\footnote{For a recent discussion of ``stellar enhancement factors", and the related but distinct ``ground state fraction of
the stellar rate", see Rauscher et al.\cite{Rau11}.}.

It is often not appreciated that under certain special conditions the thermonuclear rate of a given reaction is irrelevant, despite the fact that the main abundance flow passes through this nuclear link. This situation may arise, for example,
at sufficiently high temperatures when the abundance flow encounters a waiting point with a very small (less than a few hundred keV), or even negative, proton separation energy (i.e., the minimum energy required to initiate proton emission). In this case, an equilibrium between the rates of the forward
reaction and corresponding reverse photodisintegration is quickly established. Clearly, once equilibrium has been established for a given pair of nuclei $A$ and $B$, the abundance flow and nucleosynthesis near the waiting point become
independent of the reaction rates linking these two species. To be more specific, consider nuclides $A$, $B$ and $C$ located near the proton dripline, where an equilibrium $A+p \leftrightarrow B+\gamma$ is established. Furthermore, assume that
species $B$ may only be depleted via another reaction, $B+p \rightarrow C+\gamma$. The nucleosynthesis depends then on the relative magnitude of the decay constants for the competing $\beta$-decay and the sequential two-proton capture of the
waiting point nucleus $A$. The latter decay constant is given by \cite{Ili07}
\begin{equation}
\lambda_{A \to B \to C} =
N_p \left(\frac{h^2}{2\pi}\right)^{3/2} \frac{1}{\left(m_{Ap}kT\right)^{3/2}} \frac{(2j_B + 1)}{(2j_A + 1)(2j_p + 1)}
\frac{G_B^{\mathrm{norm}}}{G_A^{\mathrm{norm}} G_p^{\mathrm{norm}}}e^{Q_{A \to B}/kT} \lambda_{B \to C}\label{equilibriumdecayratio}
\end{equation}
where $N_p$ is the proton number density, $h$ is Planck's constant, $m_{Ap}$ is the reduced mass of $A+p$, the $j_i$ are the nuclear spins, and the $G_i^{norm}$ are the normalized partition functions (i.e., a factor expressing the thermal distribution of nuclear levels in a given nuclide). The last two factors in Eq.~(3) contain the most important quantities: the
decay constant for sequential two-proton capture depends on the Q-value (i.e., the nuclear energy released), but not the reaction rate, of the $A(p,\gamma)B$ reaction and, in addition, on the reaction rate of the {\it subsequent} $B(p,\gamma)C$ reaction. Because the
Q-value enters exponentially in the above expression it must be known with good precision, say, better than a few tens of keV \cite{Par09}. Particularly important are the Q-values for the (p,$\gamma$) reactions on the waiting point nuclei
$^{64}$Ge, $^{68}$Se and $^{72}$Kr \cite{Ili07}. Many of the Q-values, or equivalently, the nuclear masses of the respective nuclides, for reactions taking place in type I X-ray burst nucleosynthesis have been measured experimentally, although
a number of nuclear masses need to be estimated by extrapolating experimental values \cite{AME03}. For several crucial links, the reaction Q-values are currently not based on measurements (see Section \ref{expeff}). Besides Q-values, the rates
of second-step reactions in sequential two-proton capture on waiting point nuclei, for example, $^{65}$As(p,$\gamma$)$^{66}$Se, $^{69}$Br(p,$\gamma$)$^{70}$Kr and $^{73}$Rb(p,$\gamma$)$^{74}$Sr, are expected to be important for nucleosynthesis
considerations (Section \ref{sens}).

Apart from Q-values and reaction rates, decay constants for $\beta$-decays are needed for modeling the nucleosynthesis in type I X-ray bursts. These have been measured with sufficient accuracy for most nuclides up to mass $A=100$ \cite{Aud03} (see
Section \ref{expeff}). As was the case for reaction rates, the ``laboratory decay constants" should generally be replaced by ``stellar decay constants" \cite{Ful82,Oda94,Lan01}. For type I XRBs the correction factors are presumably small
for most nuclides, though detailed calculations are needed to better quantify these factors and examine their effects in XRB models.

\subsection{Overview and applicability of statistical model calculations}
\label{secHF}

As discussed in Section \ref{modpred}, a typical type I XRB reaction network calculation contains thousands of nuclear links (reactions and weak interactions). Only a tiny fraction of the thermonuclear reaction rates are derived from directly measured nuclear physics input. The
few direct measurements that have been performed, necessarily involving radioactive ion beams, focused mainly on the breakout sequences from the CNO mass range (Section \ref{expeff}). It is thus clear that the vast majority of reaction rates
must be adopted from a nuclear reaction model.

The most successful model for theoretically estimating a large number of reaction rates for nucleosynthesis studies is based on the Hauser-Feshbach theory of nuclear reactions \cite{Hau52,Vog68}. It assumes that near the incident energy there
is a large number of levels for each quantum number (spin and parity) in the compound nucleus through which the reaction can proceed, implying a large level density in the astrophysically important excitation energy range of the
compound nucleus. If this condition is fulfilled then the Hauser-Feshbach theory requires very few ingredients for predicting cross sections or reaction rates. The most important ingredients are transmission coefficients and level densities.
Both of these input quantities could be fine-tuned for a given reaction of interest, for example, by adopting parameters that reproduce measured cross sections. In reality, however, the number of unmeasured reactions is very large and it is
necessary to compute the desired cross sections and reaction rates with {\it global} instead of {\it local} parameters. For comparisons of Hauser-Feshbach predictions with measured cross sections, see Refs. \cite{Rau97,Sar82,Arn03}. 

It is
commonly assumed in nuclear astrophysics that the Hauser-Feshbach model can reliably predict reaction rates on average within a factor of 2 (e.g., Refs. \cite{Hof99,Rau00}, but see below). Another major advantage of the Hauser-Feshbach model is
that the effects of thermal target excitations on the reaction rate can be included in a straightforward manner. An extensive discussion of the Hauser-Feshbach model in nuclear astrophysics applications can be found in Goriely et al.\cite{Gor08} and Rauscher \cite{Rau11}.

\begin{figure}
\begin{center}
\epsfxsize=6in {\epsfbox{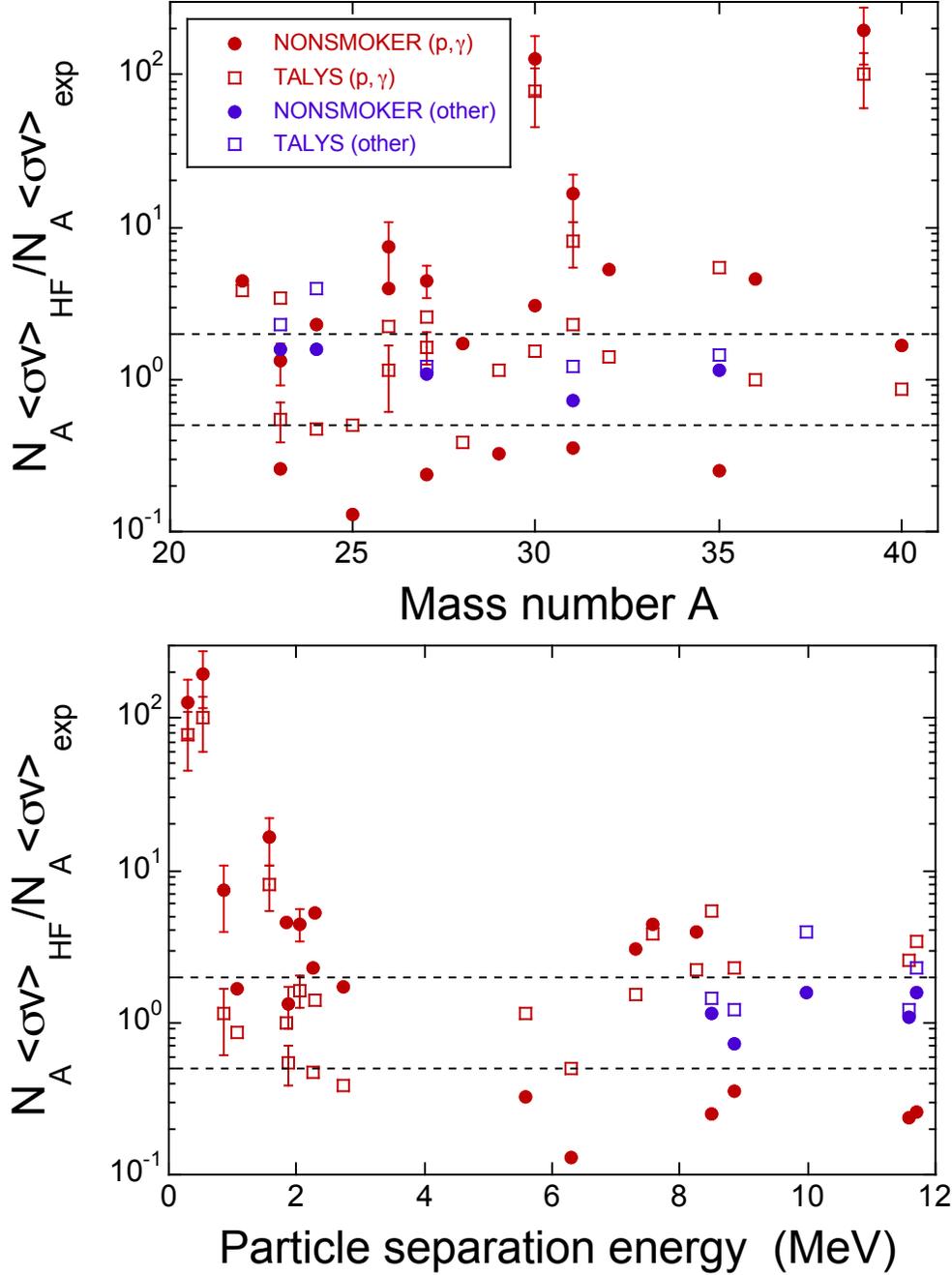}}  
\caption{Comparison of theoretical (Hauser-Feshbach) and experimental ``laboratory" reaction rates. The theoretical rates are obtained using the codes NON-SMOKER (filled circles\cite{Rau00}) and TALYS (open
squares; courtesy of S. Goriely (2012)), while the experimental results (including uncertainties) are adopted from a recent evaluation \cite{Ili10}. Ratios for (p,$\gamma$) reactions are shown in red online, while those for
(p,$\alpha$) or ($\alpha$,$\gamma$) reactions are displayed in blue online. The rates are compared at the highest temperature (different for each reaction) at which the experimental reaction rates can be fully based on nuclear physics data (i.e., at
$T_{match}$, in the parlance of Ref. \cite{Ili10}). (Top) Reaction rate ratios versus mass number of the target nucleus. (Bottom) Reaction rate ratio versus projectile separation energy (equal to the Q-value for particle capture). The
horizontal dashed lines represent ratios of a factor of 2 up or down. Note the large number of values, for both computer codes, that fall outside the commonly presumed ``factor of 2 average reliability" of Hauser-Feshbach reaction rates
\cite{Hof99,Rau00}.}
\label{HF}
\end{center}
\end{figure}

How well do current Hauser-Feshbach rates for charged-particle reactions reproduce the rates based on experiment? A preliminary answer is given in Fig.~\ref{HF}, showing the ratios of theoretical and experimental rates
versus both mass number (top panel) and projectile separation energy (bottom panel). The experimental rates, including statistically rigorous uncertainties, are adopted from a recent Monte Carlo evaluation \cite{Ili10}. Rates calculated using the
code NON-SMOKER \cite{Rau00} are shown as filled circles, while those computed with the code TALYS \cite{TALYS} are displayed as open squares. Rates for proton-capture are shown in red online and those for (p,$\alpha$) or ($\alpha,\gamma$)
reactions in blue online. For each reaction the theoretical and experimental rates are compared at the highest temperature ($T_{match}$, different for each reaction) at which the experimental rates can be fully based on nuclear physics data.
For all ratios shown, the ``minimum temperature for applicability of the Hauser-Feshbach model" ($T_{min}$, as discussed in Ref. \cite{Rau97}), is smaller than $T_{match}$. A few interesting observations can be
made:

\begin{itemize}
\item for (p,$\gamma$) reactions on target nuclei in the $A=20-40$ range, there is considerable scatter in the ratios of Hauser-Feshbach and experimental reaction rates (note the dashed horizontal lines, indicating a factor of 2 deviation), up
to factors of 100 in a few cases.
\item the theoretical (p,$\gamma$) rates from NON-SMOKER and TALYS also differ considerably in several instances, up to an order of magnitude.  This is not surprising, considering that these codes employ different global models for the nuclear physics input (i.e., transmission coefficients and level densities). 
\item for the five (p,$\alpha$) or ($\alpha,\gamma$) reactions shown in blue online, the agreement between Hauser-Feshbach and experimental results, as well as between NON-SMOKER and TALYS rates, is much better than for (p,$\gamma$) reactions.
This is especially surprising because relatively fewer levels will contribute to the rates of (p,$\alpha$) and ($\alpha$,$\gamma$) reactions compared to proton capture because of angular momentum selection rules (at the same excitation energy in the compound nucleus above
the projectile separation energy).
\item for projectile separation energies below 3 MeV, about half of the rate ratios exceed the ``factor of 2" deviation band; i.e., there is a large risk that adopting Hauser-Feshbach results will overestimate the actual rate significantly. This
is not surprising because the level densities at these excitation energies are relatively small and the conditions for the applicability of statistical reaction models are rarely fulfilled. However, this also implies that the minimum temperature,
$T_{min}$, for applicability of the Hauser-Feshbach rates \cite{Rau97}) is significantly underestimated.
\item for projectile separation energies above 5 MeV there is generally better agreement, as expected from level density arguments. There is some scatter, but {\it on average} the Hauser-Feshbach calculations seem to reproduce the experimental
rates within a factor of two.
\end{itemize}

It should be clear that Fig.~\ref{HF} does not represent a rigorous test of the reliability of Hauser-Feshbach reaction rates 
for type I X-ray burst nucleosynthesis simulations. First, the data set is rather small, especially for the (p,$\alpha$) and ($\alpha,\gamma$) reactions shown in blue online. Second, nothing can be concluded regarding other types of reactions, such as
($\alpha$,p), or for target masses in excess of $A=40$. Third, for obvious reasons, the ratios presented here apply to stable target nuclei only. However, one may suspect that the deviations between Hauser-Feshbach and experimental rates will
be larger for unstable target nuclei. Clearly, more work is needed to quantify the effects discussed above.

\subsection{Sensitivity studies to address nuclear physics needs}
\label{sens}

We have mentioned the large number of nuclides and interactions needed to accurately determine the nucleosynthesis in type I X-ray bursts, whether within the framework of multi-zone hydrodynamic simulations or simpler one-zone models. Sufficient experimental information to calculate an experimentally-based thermonuclear rate over XRB temperatures only exists for a limited number of reactions. Therefore, most reaction rates in an XRB model
depend upon theoretical calculations, as discussed in Section \ref{secHF}. To reduce the reliance on these theoretical rates, experimentalists
should measure nuclear physics quantities (e.g., resonance energies, resonance strengths) needed to extend existing experimentally-based rates to the relevant temperatures for XRBs. Furthermore, reaction rate equilibria can be better quantified and the input to theoretical rate calculations can be
improved through measurements of, e.g., nuclear masses. Since nucleosynthesis in XRBs eventually extends far into the proton-rich side beyond the valley of stability (Section \ref{modpred}), it is clear that radioactive beams are needed to perform many of the
desired measurements. To maximize the impact of these measurements, experiments should focus upon the reduction of those uncertainties in reaction rates and nuclear masses that most
significantly influence model predictions of XRB observables.
    
A useful technique to identify important nuclear interactions involves the examination of the impact on model predictions from the systematic variation of each interaction in a network by its uncertainty. Each rate may be varied
independently (requiring, in theory, as many model calculations as there are rates in the associated network) or simultaneously through a Monte Carlo method (from which e.g., correlations between an enhancement factor applied to a rate X and
the corresponding change in the yield of an isotope Y may be deduced). For theoretical rates in the network, suitable variation factors may be adopted to account for possible discrepancies, as discussed in Section \ref{secHF}. These types of
sensitivity studies have already been performed for several other astrophysical scenarios, exposing nuclear processes that affect nucleosynthesis in, e.g., classical nova explosions \cite{Ili02,Hix}, AGB stars \cite{Izz}, core-collapse supernovae \cite{The, Rap}, or massive stars \cite{Ili11}.

For type I XRBs, comprehensive sensitivity studies using full reaction networks coupled to multi-zone hydrodynamic models are not yet feasible due to computational limitations. Some progress has been made in this context through the use of different rate libraries \cite{Cyb10}, variation of groups of rates \cite{Woo04,Thi01}, or the variation of a few specific rates \cite{Fis04,Fis06,Fis07,Dav11}. For example, Cyburt et al. \cite{Cyb10} found modest changes
in the computed light curves (by $\approx$ 5\% in relative luminosity) and larger changes in the composition of burst ashes (greater than a factor of 2 in some cases) when exploring the effects of using two different rate libraries. Woosley et al.\cite{Woo04}
investigated the sensitivity of their calculated light curve to five sets of weak interaction rates, finding large deviations in rise times, peak luminosities $L_{peak}$ and evolution of the light curves after $L_{peak}$. Thielemann et al.\cite{Thi01} reported dramatic differences in light curves (shapes and $L_{peak}$) calculated with two networks identical except for four rates (proton-capture on $^{27}$Si, $^{31}$S, $^{35}$Ar and $^{38}$Ca). Fisker
et al.\cite{Fis04} multiplied the rates of the $^{30}$S($\alpha$,p) and $^{34}$Ar($\alpha$,p) reactions by factors of 100, finding minor effects from each variation relevant to a double-peak structure in their light curves (see Section \ref{modresults}). Finally, Fisker et al.\cite{Fis06,Fis07} and Davids et al.\cite{Dav11}, using different codes, varied the $^{15}$O($\alpha$,$\gamma$) rate within uncertainties to ascertain its impact on light curves. Curiously, contradictory results indicative of model dependencies are seen. When Fisker et al. \cite{Fis06} used what they refer to as a ``lower limit" for
the $^{15}$O($\alpha$,$\gamma$) rate, their model revealed non-bursting behaviour -- more precisely, a slowly oscillating luminosity with a period of about four hours. The use of a larger rate produced bursts in their model, with $L_{peak}$ about two orders of magnitude
greater than luminosities they found with the lower rate. On the other hand, when Davids et al.\cite{Dav11} used a very similar rate to Fisker et al.'s ``lower limit" (within a model with similar accretion rate as well) their simulation not only produced
bursts, but the \emph{brightest} bursts; namely, a \emph{larger} $^{15}$O($\alpha$,$\gamma$) rate resulted in bursts with $\approx$ 50\% \emph{lower} $L_{peak}$ values. It is clear that improved tests with different models
must be performed to clarify these discrepancies and others previously identified (e.g., the role of compositional inertia -- see Section \ref{modresults}).

\begin{figure} 
\begin{center}  
\epsfxsize=6in {\epsfbox{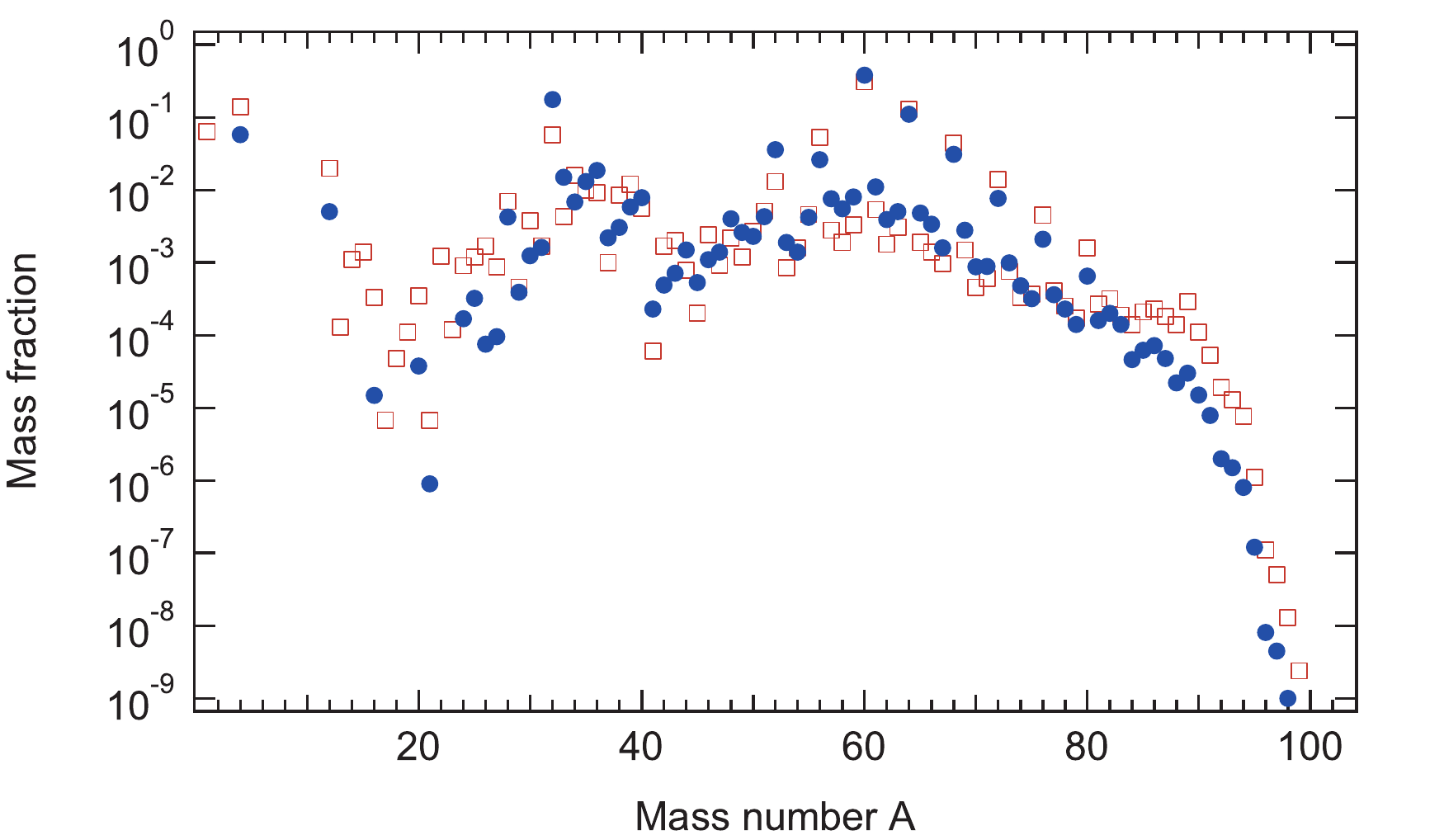}}  
\caption{Comparison of final yields from a hydrodynamic XRB model (open squares, burst 3 from model 1 of Ref. \cite{Jos10}) and from a post-processing calculation (filled circles) using the hottest shell of the same model.}  
\label{comp_hydro_pp_yields}  
\end{center}  
\end{figure}

\begin{figure}  
\begin{center}  
\epsfxsize=6in {\epsfbox{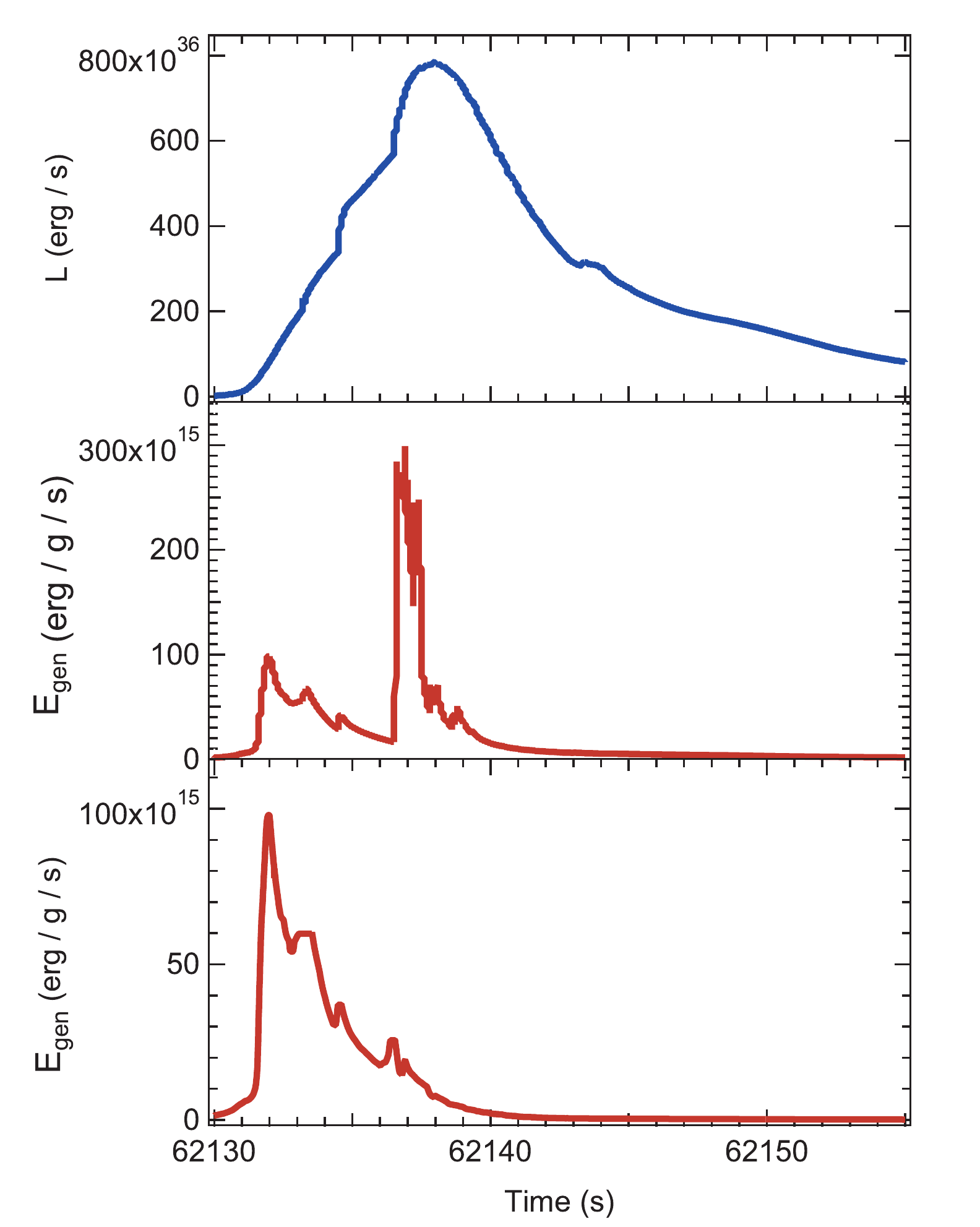}}  
\caption{(Top) Light curve from a hydrodynamic XRB model (burst 3 from model 1 of Ref. \cite{Jos10}). (Middle) Nuclear energy generation rate from the hottest shell of the same hydrodynamic model.  (Bottom) Nuclear energy generation rate from a post-processing calculation using temperature and density profiles (Fig.~\ref{profiles}) from the same shell of the same hydrodynamic model.  These plots show that the rise in luminosity is due to the nuclear energy released within the ignition shell, deep inside the envelope,
which is later convectively transported towards the surface.  The huge spike in the nuclear energy generation from the hydrodynamic model (middle panel) is due to the sudden injection of protons in the vicinity of the ignition shell following the onset of convection.  Such an effect cannot be handled within one-zone post-processing calculations (bottom panel).}
\label{comp_light_curves}  
\end{center}  
\end{figure}

Systematic sensitivity studies to guide future experimental nuclear physics measurements for XRBs have only been made with one-zone or post-processing models. As discussed in Section \ref{modover}, these are significantly less computationally expensive than multi-zone models, but, by construction,
neglect crucial hydrodynamic effects such as convection and the finite diffusion time for energy to escape from the accreted envelope. Some effects of these simplified approaches on predictions of the nucleosynthesis and energy generation are
illustrated in Figs.~\ref{comp_hydro_pp_yields} and \ref{comp_light_curves}. We emphasize that the comparisons in Figs.~\ref{comp_hydro_pp_yields} and \ref{comp_light_curves} are only strictly valid for the models discussed below.

Fig.~\ref{comp_hydro_pp_yields} shows the final mass-averaged yields from a multi-zone hydrodynamic model (open squares, burst 3 from model 1 of Jos\'e et al.\cite{Jos10}). These are compared to the final yields from a post-processing calculation using
the thermodynamic history (see Fig.~\ref{profiles}) from only the hottest shell of the same hydrodynamic model (filled circles). The abundances from the hydrodynamic model at T = 0.2 GK during the initial stages of the burst were used as initial
abundances for the post-processing calculation, and the final yields for both calculations were extracted at T = 0.2 GK during the decline of the burst.  Discrepancies are clear, particularly below A $\approx$ 25; nonetheless, at higher masses the results for both calculations show
similar trends, especially in the higher relative abundances at the waiting points (A = 60, 64, 68, 72, 76, 80).

Fig.~\ref{comp_light_curves} shows the light curve from the same hydrodynamic model mentioned above, the nuclear energy generation rate within the hottest shell of the same model, and the nuclear energy generation rate calculated during the
post-processing of the thermodynamic history (see Fig.~\ref{profiles}) from the same shell of the same hydrodynamic model. The energy generation maxima occur roughly where the waiting-point abundances change rapidly -- see e.g., Ref. \cite{Ili99}. Although one may be able to compare selected characteristic features in the nuclear energy generation curves, it is evident that any careful study of changes to the nuclear energy generation rate (let alone a predicted
light curve) owing to nuclear physics uncertainties can only be made with a full hydrodynamic model. The energy contributions from different shells as well as the time required for energy to leave the envelope clearly make non-negligible
contributions to the light curve.  As well, we stress again that convection is not included in these post-processing calculations.  

Nonetheless, if a post-processing sensitivity study relies upon a large number of thermodynamic histories, representing the available parameter spaces of temperatures and densities achieved in an XRB, the results serve as a useful guide for
future measurements. This is especially true if variations in a particular reaction rate or Q-value affect predicted quantities in multiple one-zone models (where each uses a different thermodynamic history). Such an approach is necessary in
light of the wide variety of observed XRB light curves, possibly indicating diverse conditions under which the thermonuclear runaway occurs. The usual note of caution with regard to these types of studies is worth stressing: since
post-processing calculations only track existing thermodynamic histories, results obtained from variations in nuclear processes that significantly affect the energy production should be interpreted carefully. Indeed, a self-consistent analysis
with a hydrodynamic code capable of suitably adjusting both the temperature and the density of the stellar envelope (in response to the change in energy generation) seems necessary to reliably treat such cases.

Some limited investigations of the impact of reaction rate uncertainties on predictions of one-zone or post-processing models have been made previously. For example, Iliadis et al. \cite{Ili99} varied independently the rates of proton-capture reactions
on $^{27}$Si, $^{31}$S, $^{35}$Ar and $^{39}$Ca and found only modest variations in final yields of $\approx$ 50\% or less for masses A = 40 -- 48, and negligible changes to the total energy generation. Schatz et al.\cite{Sch98} and Koike et al.\cite{Koi99} examined the impact on yields and energy
production from using different reaction rate libraries. Wallace and Woosley \cite{WW81} and Koike et al.\cite{Koi04} gauged the impact of the size of their networks on energy generation and the extent of the nucleosynthesis. Indeed, Koike et al.\cite{Koi04} found
that XRBs may produce species up to $^{126}$Xe, beyond the SnSbTe cycle; they ascribe the necessary abundance flows at the heaviest masses to their larger network compared to that of Ref. \cite{Sch01}. Clement et al.\cite{Cle03} reported considerable
differences in light curve shapes resulting from calculations with different nuclear mass models (for nuclei A = 60 -- 75); these are essentially effects due to different separation energies around waiting point nuclei, as discussed in Section
\ref{nucover}. As well, several one-zone calculations have recently been performed to examine the impact of particular new experimental results (see Section \ref{expeff}).

\begin{figure}  
\begin{center}  
\epsfxsize=6in {\epsfbox{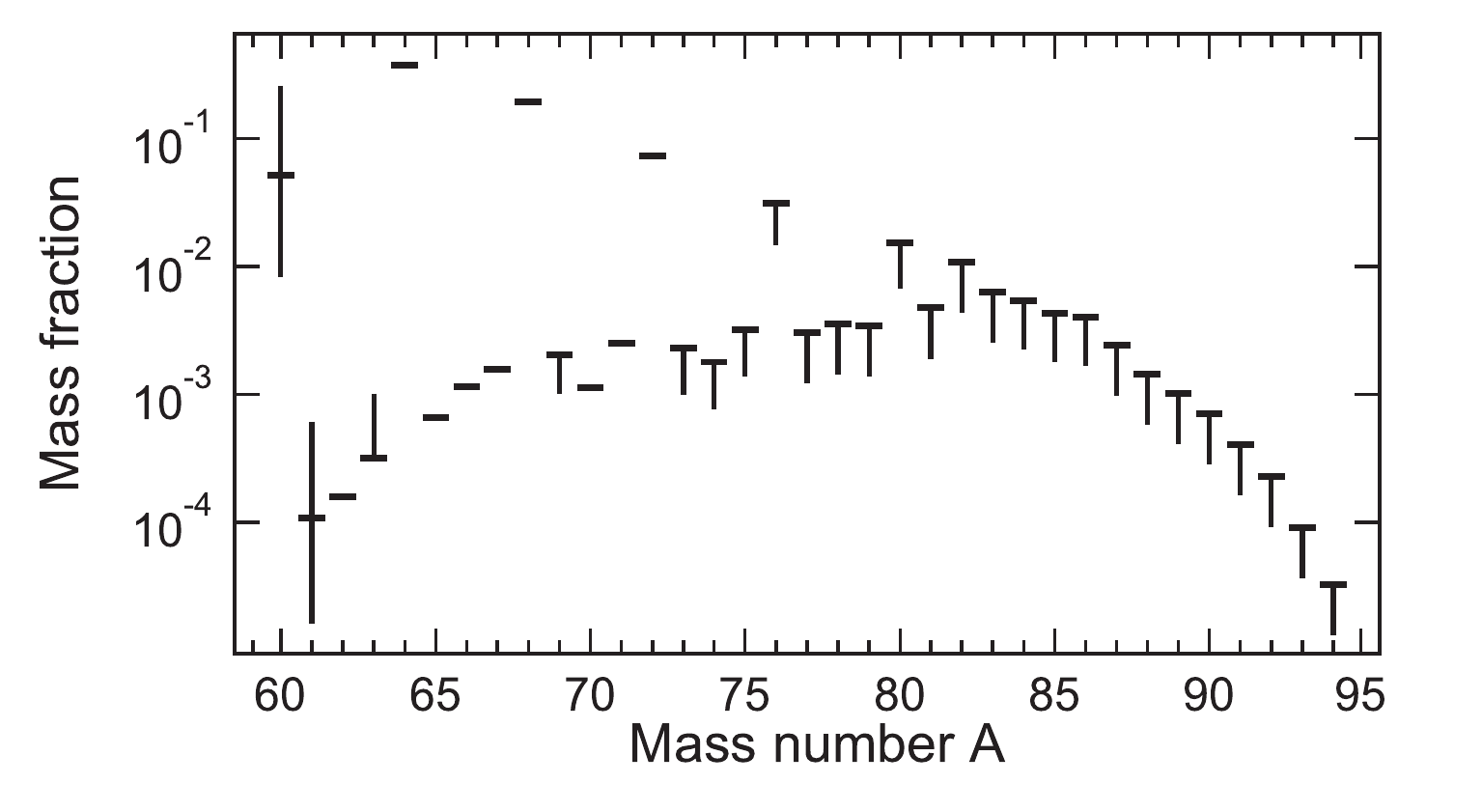}}  
\caption{The effect on final XRB yields from varying the thermonuclear rate of the $^{61}$Ga(p,$\gamma$)$^{62}$Ge reaction by a factor of 10 (up and down). Only theoretical calculations exist for this rate. The horizontal bars indicate mass
fractions using the standard rate, and the vertical error bars indicate changes to these mass fractions due to the rate variations. From post-processing calculations with model K04--B6 in Ref. \cite{Par08}.}
\label{61Gapg}  
\end{center}  
\end{figure}

\begin{figure}  
\begin{center}  
\epsfxsize=6in {\epsfbox{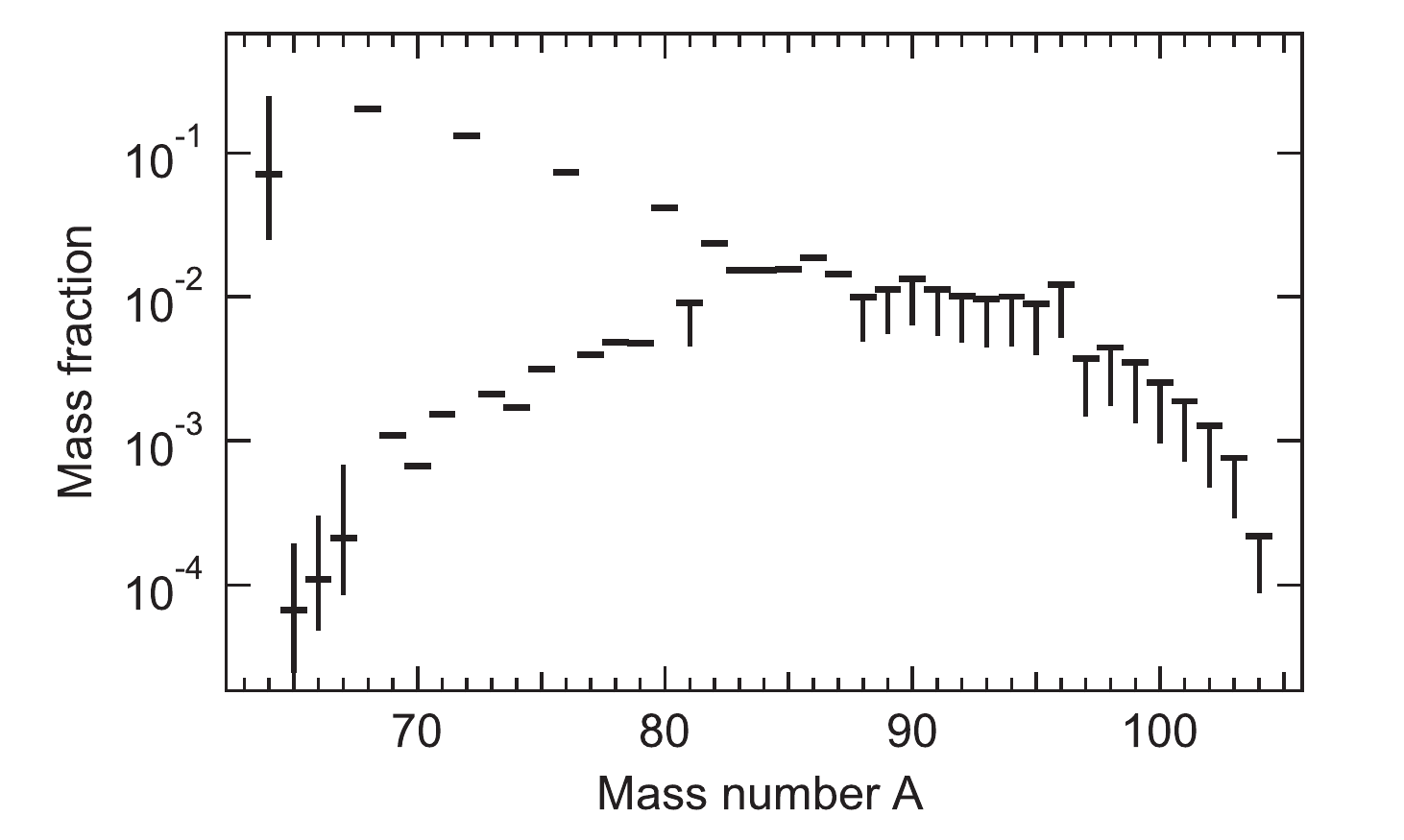}}  
\caption{Same as Fig.~\ref{61Gapg} but from varying the thermonuclear rate of the $^{65}$As(p,$\gamma$)$^{66}$Se reaction by a factor of 10 (up and down). From post-processing calculations with model K04 in Ref. \cite{Par08}. }
\label{65Aspg}  
\end{center}  
\end{figure}

The only full sensitivity studies for type I XRBs were reported in Parikh et al.\cite{Par08,Par09}. Post-processing calculations were performed using seven different thermodynamic histories, comprising models extracted from the literature and
parameterized models. These scenarios were chosen to explore the range of temperatures and densities achieved during XRBs and find common rates and reaction Q-values whose uncertainties needed to be reduced. As well, one of these histories was
tracked assuming initial abundances of different metallicities to explore the impact of this variable. Three different approaches were employed: first, the rates of all nuclear processes within the network were independently varied within
uncertainties, for each scenario; second, rates for all reactions with Q-values less than 1 MeV were recalculated assuming Q-value variations within uncertainites, and the impact of each of these variations was examined for each scenario; and
third, rates of all nuclear processes were varied simultaneously using a Monte Carlo procedure, for two scenarios. An uncertainty of a factor of 10 was assumed for all purely theoretical rates in the network. In the independent-variation study, uncertainties
in relatively few nuclear processes were observed to affect the final nucleosynthesis significantly (i.e., to change the yields of species with final mass fractions $>$ 10$^{-5}$ by at least a factor of two). The impact of variations in the $^{61}$Ga(p,$\gamma$) and $^{65}$As(p,$\gamma$)
rates is illustrated in Figs.~\ref{61Gapg} and \ref{65Aspg}. These are two examples of rates that did significantly modify the nucleosynthesis in most of the adopted scenarios; only theoretical calculations are available for both of these
rates. The Q-value variation study examined the impact of uncertainties in nuclear masses on abundance flows about nuclear waiting points. A principal result of this part of the study was the need to determine experimentally the proton separation energy $S_{p}$ of $^{65}$As (see Section
\ref{expeff}). Figure~\ref{64Gepg} (thin grey bars) shows the dramatic effect on final yields found from varying $S_{p}(^{65}$As) within the uncertainty estimated in Ref. \cite{AME03}. Finally, the results of the Monte Carlo and independent-variation approaches were in agreement: similar reactions were identified with rate uncertainties that significantly affected final XRB yields.

Given the recent availability of a new set of hydrodynamic models \cite{Jos10}, we have repeated the independent-variation sensitivity study using a thermodynamic history from one of these bursts (see Fig.~\ref{profiles}). In general, the
results are in accord with those reported in Ref. \cite{Par08}. Variations in some reaction rates (e.g., $^{61}$Ga(p,$\gamma$) and $^{65}$As(p,$\gamma$)) have similar effects on extended mass regions (as in e.g., Figs.~\ref{61Gapg} and \ref{65Aspg}). In addition,
variations in the rates of alpha-capture reactions on $^{24}$Mg, $^{28}$Si, $^{29}$Si and $^{32}$S and the $^{30}$P(p,$\gamma$)$^{31}$S reaction were found to affect the final yields of at least three isotopes around $A\approx30$ by at least a factor of two (similar to results seen with
scenarios F08, K04--B2, K04--B4, and K04--B5 in Ref. \cite{Par08}).

We note in passing that variations in the $^{15}$O($\alpha$,$\gamma$), $^{27}$Si(p,$\gamma$), $^{31}$S(p,$\gamma$) and $^{35}$Ar(p,$\gamma$) rates by experimental uncertainties did indeed affect the nuclear energy generation in the post-processing calculations of Ref. \cite{Par08} (and in the new
set of calculations described above). As such, hydrodynamic tests seem required to examine the impact of uncertainties in these rates. As mentioned above, several studies of the $^{15}$O($\alpha$,$\gamma$) rate uncertainty have already been made, but new hydrodynamic calculations of
the impact of uncertainties in the other rates seem warranted given the results of Refs. \cite{Ili99,Thi01}, as well as new experimental results that may affect these rates \cite{Wre10b}.

As full sensitivity studies with multi-zone hydrodynamic models are not yet possible, and one-zone or post-processing calculations suffer from necessary simplifications, an obvious compromise would be to first identify reactions of importance
through the latter approaches and then investigate variations in only those selected reactions through hydrodynamic models. Such studies are in progress \cite{Amt06,Rob06}, however initial results indicate that this method may
not be a satisfactory substitute for a full sensitivity study using only hydrodynamic models. For example, Amthor et al. \cite{Amt06} found conflicting results when performing rate variations in one-zone and hydrodynamic models: variation of the $^{46}$Cr(p,$\gamma$) rate produced similar results with the two approaches, but differences in the results appeared when the $^{42}$Ti(p,$\gamma$) or $^{49}$Fe(p,$\gamma$) rates were varied. As well, hydrodynamic tests with the processes identified in Refs. \cite{Par08,Par09} are already underway.

\subsection{Recent experimental efforts to improve model predictions}
\label{expeff}

\begin{figure}  
\begin{center}  
\epsfxsize=6in {\epsfbox{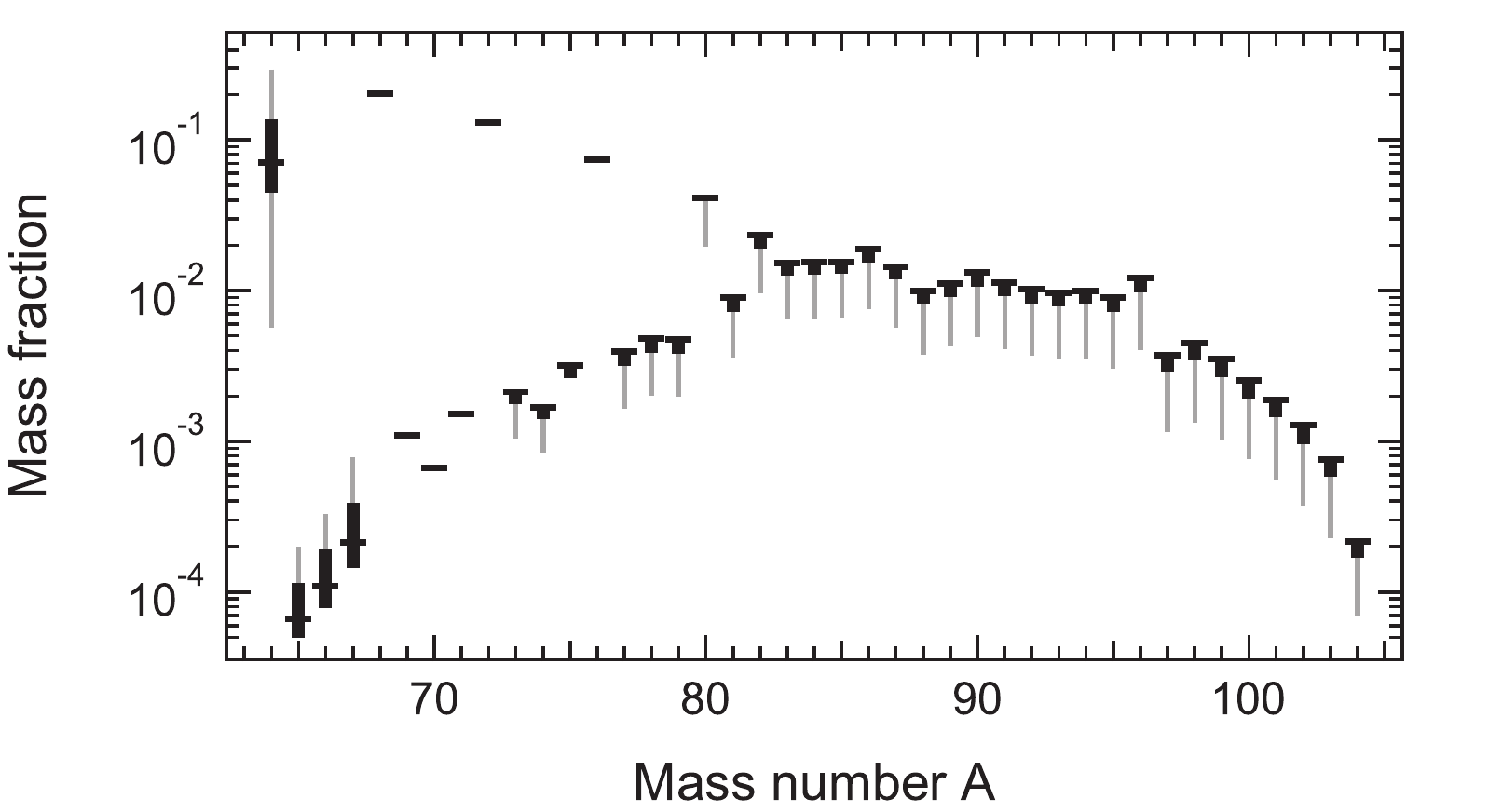}}  
\caption{The effect on final XRB yields from varying the reaction Q-value of the $^{64}$Ge(p,$\gamma$)$^{65}$As reaction within uncertainties. The horizontal bars indicate mass fractions using either $Q_{AME03} = -80 \pm 300$ keV or $Q_{exp} =-90 \pm 85$ keV (the former
uses the estimated value and uncertainty from Ref. \cite{AME03}, while the latter uses experimental values for all masses \cite{Sch07,Tu11}. The vertical error bars indicate the changes to these mass fractions due to the uncertainty in $Q_{AME03}$
(thin grey bars) or the uncertainty in $Q_{exp}$ (thick black bars). From post-processing calculations with model K04 \cite{Par08,Par09}.}
\label{64Gepg}  
\end{center}  
\end{figure}

A number of nuclear physics experiments have been performed in recent years to better define model predictions of nucleosynthesis and light curves for type I XRBs. The majority of these are high-precision mass measurements, highlighted by the determinations of
masses to experimentally constrain the proton separation energies of $^{65}$As and $^{69}$Br for the first time ($-90 \pm 85$ keV and $-785^{+34}_{-40}$ keV respectively \cite{Sav09,Rog11,Tu11,Sch07}, to be compared
to the estimated values and uncertainties of $-80 \pm 300$ keV and $-450 \pm 100$ keV \cite{AME03}). The impact of the reduction in the uncertainty of $S_{p}(^{65}$As) was explored through a one-zone model \cite{Tu11}, where the precision of light curve and nucleosynthesis
predictions was seen to be improved. This is also seen in Fig.~\ref{64Gepg}, where the thick black error bars indicate the variation in final yields from post-processing calculations in which $S_{p}$($^{65}$As) was varied within its experimental
uncertainty. These
 results are in agreement with those presented in Ref. \cite{Tu11}: only immediately about A = 64 are the predictions still sensitive to the uncertainty in $S_{p}$($^{65}$As) (and even these yields change by less than a factor of two). This is a
key issue when considering the required precision of a mass value to constrain nucleosynthesis predictions: the answer clearly depends upon which mass is being measured and which mass region is being examined in the resulting XRB
yields. This has been addressed by Schatz et al.\cite{Sch06} in the context of changes to the effective lifetimes for $^{64}$Ge, $^{68}$Se and $^{72}$Kr from uncertainties in nuclear masses, and in the calculations of Parikh et al.\cite{Par09} where variations of a few
specific reaction Q-values were made by then-current uncertainties and 20\% of those uncertainties. For example, it was found that varying $S_{p}$($^{64}$Ge) by $\pm$ 300 keV affected a large number of yields (thin grey error bars in Fig.~\ref{64Gepg})
while variations by 0.2$\times$300 = 60 keV did not change any yields by more than a factor two.

The expected effect of the measured $S_{p}(^{65}$As) and $S_{p}(^{69}$Br) values would be a reduction in the abundance flow beyond A $\approx$ 70 \cite{Rog11} relative to predictions using the estimated masses from Ref. \cite{AME03}. This is supported by preliminary hydrodynamic calculations using model 1 from Ref. \cite{Jos10}, although the influence of the new measurements on the mean composition of the envelope following a burst is not large.  Indeed, we find reductions in mass fractions of less than 30\% for nuclides A $>$ 65.  Additional hydrodynamic models are needed to explore further the precise impact of these measurements in detail. 

Other recent relevant mass measurements have provided precise values for masses A $>$ 80, including proton-rich isotopes of elements between Rb and I \cite{Kan06,Web08,Bre09,Elo09a,Elo09b,Hae11,Fal11}\footnote{See also databases at http://www.nuclearmasses.org/ \\http://research.jyu.fi/igisol/JYFLTRAP\_masses/ \\http://www-win.gsi.de/aptrap/database/index.asp \\http://isoltrap.web.cern.ch/isoltrap/database/isodb.asp}.
Some of these studies have been motivated in part by exploring the still-unclear maximum endpoint of nucleosynthesis in XRBs \cite{Elo09a} or explaining the solar abundances of the light p-nuclei $^{92,94}$Mo and $^{96,98}$Ru through contributions from XRBs (see Section \ref{modresults}). At lower masses, measurements have improved the
$S_{p}$($^{57}$Cu) value \cite{Kan10} and the $^{27}$Si(p,$\gamma$) rate through a new measurement of the $^{28}$P mass \cite{Wre10b}. From the list of masses desired to constrain nucleosynthesis in the post-processing calculations of Ref. \cite{Par08,Par09}, experimental determinations of the masses of $^{26}$P, $^{27}$S, $^{43}$V, $^{46}$Mn, $^{47}$Mn, $^{51}$Co, $^{56}$Cu, $^{62}$Ge, $^{66}$Se, $^{70}$Kr, $^{84}$Nb, $^{86}$Tc, $^{89}$Ru, $^{90}$Rh,
$^{96}$Ag, $^{97}$Cd, $^{99}$In and $^{103}$Sn are still needed. In addition, the experimentally-known masses of $^{31}$Cl, $^{45}$Cr, $^{61}$Ga,$^{71}$Br and $^{83}$Nb should be determined to better precision.  These desired mass measurements are summarized in Table~\ref{tablemasses}.  

With regard to lifetimes of radioactive nuclei involved in XRB nucleosynthesis, the required $\beta^{+}$ decay and electron capture rates depend upon the temperature and density of the environment.  While such ``stellar" weak interaction rates have been computed for different ranges of nuclei (e.g., A = 1 -- 39 \cite{Oda94}; A = 45 -- 65 \cite{Lan01}; both using the shell model) no single consistent treatment of stellar lifetimes for all isotopes in an XRB nucleosynthesis network exists.  Although the differences between ``laboratory" and ``stellar" lifetimes are presumably small given the temperatures involved in XRBs (certainly compared to supernovae) the effects of these differences on model predictions need to be examined in more detail.  We note that the vast majority of ``laboratory" lifetimes for the nuclei of interest in XRBs have not only been measured, but indeed, are claimed with uncertainties of less than $\approx$30\% \cite{Aud03}.  To illustrate, even for the species with unknown masses listed above, only the lifetime of $^{51}$Co is unknown; lifetimes of the other nuclei are all known to $\approx$ 30\%.  Some of these values may however be based upon only one measurement and so efforts to ensure the reliability of these evaluated lifetimes are welcome, especially at waiting points \cite{Baz08,Sto09}.  Present experimental uncertainties in ``laboratory" lifetimes (i.e., $\pm$30\% or better) were found to affect significantly neither any XRB yields nor the nuclear energy generation in the post-processing studies of Ref. \cite{Par08}.

A few direct or indirect experiments have been performed recently to improve several reaction rates for XRB calculations, including the breakout reactions $^{14}$O($\alpha$,p) \cite{He09}, $^{15}$O($\alpha$,$\gamma$) \cite{Tan07}, $^{18}$Ne($\alpha$,p) \cite{Mat09,Sal12}, and $^{21}$Na(p,$\gamma$) \cite{Dau04} and reactions around the $^{30}$S waiting point, such as $^{29}$P(p,$\gamma$) \cite{Gal10,Set10, Set11}, $^{30}$P(p,$\gamma$) \cite{Wre09a}, $^{30}$S(p,$\gamma$) \cite{Wre09b}, $^{30}$S($\alpha$,p) \cite{Dei11}, $^{32}$Cl(p,$\gamma$) \cite{Sch05}, and $^{35}$Ar(p,$\gamma$) \cite{Wre10a}.

\section{Outlook} 
\label{outlook}

Tremendous progress has been made in recent years to better understand nucleosynthesis in type I X-ray bursts and the role of nuclear physics uncertainties in model predictions of observables.  Prospects for stronger observational tests of XRB nucleosynthesis predictions depend upon future X-ray missions and the development of improved models for the atmospheres of neutron stars.  The Rossi X-ray Timing Explorer had been the most versatile observatory for XRBs over the past decade because of its ability to monitor the sky for transient outbursts from low-mass X-ray binaries and subsequently trigger pointed observations with its higher resolution instruments.  Fortunately, other missions still in orbit have wide field cameras onboard (Fermi\cite{Fermi}, Swift/BAT\cite{Swift}, MAXI\cite{MAXI}, and soon, eROSITA\cite{eROSITA}) while detailed pointed measurements are available with XMM-Newton\cite{XMM}, Chandra\cite{CHANDRA}, Swift/XRT or Suzaku\cite{Suzaku}.  Some recent missions such as NuSTAR \cite{NuSTAR} (launched in June 2012) and future, proposed projects such as the Large Observatory For X-ray Timing (LOFT \cite{LOFT}) will help to advance our understanding of XRBs.  The Large Area Detector planned for LOFT would have an effective area of $\approx$12 m$^{2}$, which is an order of magnitude larger than current spaceborne X-ray detectors (in the 2 -- 30 keV energy range).  This instrument, as well as NuSTAR \cite{chen09}, should be able to detect absorption edges in PRE bursts.  The second instrument onboard LOFT, the Wide Field Monitor, would monitor transient sources over a large fraction of the sky.  With these instruments, LOFT would be able to probe the spectral variability of sources down to unprecedentedly short timescales.  Unfortunately, no currently planned mission, LOFT included, nor any mission currently in orbit has both the collecting area and the necessary energy resolution to measure absorption lines in neutron star atmospheres over reasonable timescales.   

Improved models of neutron star atmospheres are needed to aid in the interpretation of the observed neutron star spectrum.  The nature of the ashes of thermonuclear burning during XRBs represents a key ingredient.  Currently, the use of atmosphere models is limited to the determination of the colour correction factor from assumed atmosphere abundances.  This factor is then used to determine the actual neutron star photospheric temperature from the deduced blackbody temperature.  Direct fitting of the observed spectra with a grid of atmosphere models could provide constraints on the abundances of species present in the neutron star atmosphere.  This approach, combined with high signal-to-noise, time-resolved spectroscopy could, in principle, even probe changes in the chemical composition of the atmosphere during a burst.   

The coupling of state-of-the-art hydrodynamic models and large nuclear reaction networks has led to a better understanding of the complex interplay between nuclear interactions and X-ray burst properties (such as light curves, burst recurrence times or $\alpha$ values).  It is expected that better agreement with observations will be achieved through the incorporation of the neutron star spin and magnetic field in models.  Moreover, since aspects such as convective transport and burning front propagation can only be accurately modeled through multi-dimensional simulations, such studies must be performed whenever the necessary resources become available. In contrast with related astrophysical scenarios such as classical novae, no mass ejection results from current hydrodynamic simulations of XRBs.  This aspect simplifies the treatment of the 
problem since purely Lagrangian schemes are not strictly needed to properly follow the expansion of 
the burst envelope. However, one should bear in mind that the modeling of both sites, novae and X-ray 
bursts, is extremely challenging because of the deflagrative nature of the burning front.  Rigorous treatments will likely require the use of low Mach number codes \cite{Mal11}, and such studies are therefore encouraged. As well, the need to rely on multiple bursts to account for the long-term evolution of XRB sources makes multi-dimensional simulations extremely challenging, and it is expected that models with spherical symmetry will persist.  Alternative ways to better constrain XRB nucleosynthesis should also be explored.  This should include the detailed modeling of the process of photospheric radius expansion, the possibility of the ejection of nuclear-processed material through radiation-driven winds and the imprint of these winds on the burst spectrum.       

Several radioactive beams facilities capable of making nuclear physics measurements needed to improve estimates of reaction rates in XRBs are online or will be online shortly (e.g., TRIUMF-ISAC\footnote{http://www.triumf.ca/}, RIKEN-RIBF\footnote{http://www.nishina.riken.jp/RIBF/}, GSI-FAIR\footnote{http://www.gsi.de/en/research/fair.htm}, GANIL-SPIRAL2\footnote{http://www.ganil-spiral2.eu/}, MSU-FRIB\footnote{http://www.frib.msu.edu/}).  Through post-processing calculations, nuclear physics uncertainties that may most significantly affect model predictions have been identified.  Measurements of the masses listed in Table~\ref{tablemasses} are needed to constrain abundance flows about waiting-point nuclei or improve theoretical rate calculations. We stress the need for nuclear structure information above the proton thresholds in $^{62}$Ge, $^{66}$Se and $^{70}$Kr to better determine the rates of the $^{61}$Ga(p,$\gamma$), $^{65}$As(p,$\gamma$) and $^{69}$Br(p,$\gamma$) reactions which act upon equilibrium abundances following waiting points. Resolving these reaction rates should be priorities for experimentalists.   In addition, measurements to further improve the rates of breakout reactions such as $^{14}$O($\alpha$,p), $^{15}$O($\alpha$,$\gamma$) or $^{18}$Ne($\alpha$,p); rates around A = 30, particularly of interest for multi-peaked structures in predicted light curves; and others, such as $^{57}$Cu(p,$\gamma$) or $^{96}$Ag(p,$\gamma$) would be welcome.  Furthermore, since ``stellar" decay constants should ideally be used in XRB models rather than ``laboratory" decay constants, we encourage the development of improved, consistent treatments for calculating stellar weak interaction rates for all isotopes in a typical XRB network.  Finally, comprehensive sensitivity studies employing hydrodynamic models are needed to directly gauge the impact of nuclear physics uncertainties on observable properties of type I X-ray bursts.

\section*{Acknowledgments}

This work has been supported by MICINN grants AYA2010-15685 and AYA2011-23102, Government of Catalonia grant 2009SGR-1002, and the ESF EUROCORES Program EuroGENESIS through the MICINN grant EUI2009-04167.  C.I. acknowledges support from the US Department of Energy under grant DE-FG02-97ER41041 and the National Science Foundation under award number AST-1008355.


\newpage

\begin{table}[ht]
  \caption{Mass measurements desired to improve calculations of nucleosynthesis in XRBs \cite{Par08,Par09}.  Estimated masses and uncertainties from Ref. \cite{AME03} are given with a \# symbol; increased precision is required for the other, experimental masses listed.  Masses required primarily to better quantify reaction rate equilibria at waiting point nuclei (W) or refine theoretical rate calculations (T) are indicated.}
  \begin{center}
    \begin{tabular}{|l|l|l|}
   \hline
    Nuclide&Mass excess\cite{AME03} (keV)&Purpose\\
   \hline
   $^{26}$P &\#$10973\pm196$&W \\
   $^{27}$S &\#$17543\pm202$&W \\
   $^{31}$Cl &$-7067\pm50$&W \\
   $^{43}$V &\#$-18024\pm233$&W \\
   $^{45}$Cr &$-18965\pm503$&W \\
   $^{46}$Mn &\#$-12370\pm112$&W \\
   $^{47}$Mn &\#$-22263\pm158$&W \\
   $^{51}$Co &\#$-27274\pm149$&W \\
   $^{56}$Cu &\#$-38601\pm140$&W \\
   $^{61}$Ga &$-47090\pm53$&W \\
   $^{62}$Ge &\#$-42243\pm140$&T \\
   $^{66}$Se &\#$-41722\pm298$&T \\
   $^{70}$Kr &\#$-41676\pm385$&T \\
   $^{71}$Br &$-57063\pm568$&T \\
   $^{83}$Nb &$-58959\pm315$&T \\
   $^{84}$Nb &\#$-61879\pm298$&T \\
   $^{86}$Tc &\#$-53207\pm298$&T \\
   $^{89}$Ru &\#$-59513\pm503$&W \\
   $^{90}$Rh &\#$-53216\pm503$&W \\
   $^{96}$Ag &\#$-64571\pm401$&T \\
   $^{97}$Cd &\#$-60603\pm401$&T \\
   $^{99}$In &\#$-61274\pm401$&W \\
   $^{103}$Sn &\#$-66974\pm298$&T \\
   \hline
  \end{tabular}
\label{tablemasses}  
  \end{center}
\end{table}

\end{document}